\documentclass[prd,showpacs,twocolumn,preprintnumbers
]{revtex4}

\usepackage{amsmath} \usepackage{graphicx} \usepackage{amsfonts}
\usepackage{array} \usepackage{amsthm}

\newcommand{\vf}{\varphi} \newcommand{\nn}{\nonumber}
\newcommand{\qd}{\qquad} \newcommand{\ra}{\rightarrow}
\newcommand{\mn}{{\mu\nu}}
\newcommand{\Lef}{\left(} \newcommand{\Rig}{\right)}
\newcommand{\p}{\partial} \newcommand{\ep}{\varepsilon}
\newcommand{\bR}{\mathbb{R}} 
\newtheorem{lem}{Lemma}
\newtheorem{theor}{Theorem}

\begin{document}
\preprint{DTP-MSU/06-20}  

\title{Cylindrically symmetric solitons in Einstein--Yang--Mills theory}
\author{
  Dmitri V. Gal'tsov}
\email{galtsov@phys.msu.ru}\affiliation{Department of Theoretical
Physics, Moscow State University, 119899, Moscow, Russia}
\author{
    Evgeny A. Davydov}
\email{eugene00@mail.ru} \affiliation{Department of Theoretical
Physics, Moscow State University, 119899, Moscow, Russia}
\pacs{04.20.Jb}
\date{00.00.00}

\begin{abstract}
Recently new Einstein-Yang-Mills (EYM) soliton solutions were
presented which describe superconducting strings with Kasner
asymptotics (hep-th/0610183). Here we study the static
cylindrically symmetric SU(2) EYM system in more detail.  The
ansatz for the gauge field corresponds to superposition of the
azimuthal $B_\varphi$ and the longitudinal $B_z$ components of the
color magnetic field. We derive sum rules relating  data on the
symmetry axis to asymptotic data and show that generic asymptotic
structure of regular solutions is Kasner. Solutions starting with
vacuum data on the axis generically are divergent. Regular
solutions correspond to some bifurcation manifold in the space of
parameters which has the low-energy limiting point corresponding
to string solutions in flat space (with the divergent total
energy) and the high-curvature point where gravity is crucial.
Some analytical results are presented for the low energy limit,
and numerical bifurcation curves are constructed in the
gravitating case. Depending on the parameters, the solution looks
like a straight string or a pair of straight and circular strings.
The existence of such  non-linear superposition of two strings
becomes possible due to self-interaction terms in the Yang-Mills
action which suppress contribution of the circular string near the
polar axis.

\end{abstract}

\maketitle
\section{Introduction}
The discovery by Bartnik and McKinnon \cite{Bartnik:1988am} of a
particle-like solution  in the Einstein-Yang-Mills (EYM) theory
and  construction of the corresponding black holes
\cite{Volkov:1989fi} stimulated wide research of
Einstein-Yang-Mills (EYM) solitons and their generalization in
supergravities and other related models  (for a review see
\cite{Volkov:1998cc}). Between them, cylindrically symmetric
four-dimensional configurations remained relatively less explored.
Useful general analysis can be found in the study of
Einstein--Maxwell geons \cite{Wheeler:1955mt,Misner:1957mt}, the
Melvin magnetic universe \cite{Melvin:1963qx} and the abelian
cosmic strings \cite{Garfinkle:1985hr}. Other configurations such
as the Yang--Mills vortex in  4+1 dimensions \cite{Volkov:2001tb},
cylindrical black holes \cite{Lemos:1994xp, Hayward:1999ek} and
some numerical solution in \cite{Slagter99} are also relevant.
Recently a new class of the EYM cylindrically-symmetric solutions
was reported \cite{GDV06} describing the Einstein-Yang-Mills
strings. The purpose of this paper is to study the cylindrically
symmetric four-dimensional EYM system in more detail.

 Our ansatz for the
gauge field contains two functions which describe the longitudinal
and the azimuthal magnetic field components. The stress-energy
tensor does not exhibit the boost symmetry as in the case of
cosmic strings, so the metric is parametrized by three different
functions of the radial variable. We analyze first the solutions
in flat space-time imposing the regularity conditions on the polar
axis. Generically, the integral curves starting with such boundary
data on the axis diverge at finite distance exhibiting the
square-root singularity of the Yang-Mills field, reminiscent of
that in the spherically symmetric case \cite{Zotov}. There exist,
however, the bifurcation curve in the parameter space which
correspond to regular solutions. The energy density does not fall
sufficiently fast at infinity, so the total energy diverges.

When gravity is switched on,  field configurations become
effectively compact and the metric is asymptotically Ricci-flat.
It is, however, not Minkowskian, but Kasner. We derive the
integral sum rules relating the boundary values of the metric
functions on the polar axis with the Kasner parameters which show
the absence of asymptotically flat configurations of non-zero
mass. We also present a (non-rigorous) proof of existence of
regular solutions with Kasner asymptotic.  Numerical integration
reveals the following typical behavior of the solutions. In the
vicinity of the polar axis the space-time is flat. With growing
distance the longitudinal magnetic field is dominant within a
certain region. The YM field there is effectively U(1), while the
metric is approximately Melvin. With the further growing distance
the azimuthal component of the YM magnetic field comes into play
and the YM non-linearity becomes manifest. In this region the
solution may be interpreted as describing the circular magnetic
string. Finally, in the asymptotic region the metric acquires the
boost symmetry in the $t-\varphi$ plane exactly in the same way as
the metric of the cosmic string is boost-symmetric in the $t-z$
plane. Moreover, the metric component $g_{zz}$ asymptotically
tends to zero, so the space-time becomes effectively $1+2$
dimensional. This effect is similar to the asymptotical
contraction of the azimuthal dimension in the Melvin solution.
Thus, our solution may be loosely interpreted as describing the
system of interacting straight and circular magnetic strings in
the four-dimensional EYM theory.

The paper is organized as follows. In Sec. II we introduce the
gauge field ansatz and discuss the symmetries of the
one-dimensional reduced action. We show the existence of two
scaling symmetries and derive the corresponding Noether currents.
In Sec. III we discuss the equations of motions and formulate the
boundary conditions of regularity on the polar axis both in the
flat space and within the full self-gravitating treatment. Using
them, we derive the sum rules relating the local values of some
combinations of the metric functions and the YM variables to the
radial integrals of quantities which admit simple physical
interpretation in terms of the energy density and principal
pressures. Sec. IV is devoted to study of the asymptotic
conditions and investigation of the existence of the desired
asymptotically vacuum solutions. Qualitative analysis of the field
equations and the results of the numerical integration are
presented in Sec. V. In conclusion we briefly formulate our
results and sketch  further prospects.
\section{The model}
\subsection{Einstein-Yang-Mills action}
We consider the pure SU(2) Einstein-Yang-Mills action
\begin{equation}\label{eq:action}
    S=\int dx^4 \sqrt{-g}\Lef-\frac{1}{16\pi G}\bR-\frac{1}{4}F_{\mn}^a F^{a\mn}\Rig
\end{equation}
with the field tensor
\begin{equation}\label{eq:F1}
    F_{\mn}^a=\p_\mu A_{\nu}^a-\p_\nu
    A_{\mu}^a+e \ep_{abc}A_{\mu}^b A_{\nu}^c.
\end{equation}
Here $e$ is the gauge coupling constant, and
\begin{equation}\label{eq:A1}
    A_\mu =\mathrm{T}_a A_{\mu}^a
\end{equation}
is the matrix-valued  gauge  potential, with the group generators
$\mathrm{T}_a$ normalized as
\begin{equation}\label{eq:T1}
\mathrm{T}_a
\mathrm{T}_b=\frac{1}{2}\ep_{abc}\mathrm{T}_c-\frac{1}{4}\delta_{ab}.
\end{equation}
Variation of the action (\ref{eq:action}) with respect to the
metric $g^{\mn}$ leads to the Einstein equations
\begin{equation}\label{eq:Einsteq1}
    R_{\mn}=8\pi G\Lef T_{\mn}-\frac{1}{2}g_{\mn}T^{\lambda}_\lambda \Rig
\end{equation}
with the stress-energy tensor
\begin{equation}\label{Eq:T1}
    T_{\mn}=-F_{\mu\sigma}^a F_{\nu}^{a\sigma}+\frac{1}{4}g_{\mn}F_{\lambda\sigma}^a
    F^{a\lambda\sigma},
\end{equation}
which is traceless:
\begin{equation}\label{eq:TT}
    T^{\lambda}_\lambda=0.
\end{equation}
Variation with respect to the gauge field $A_{\mu}^a$ gives the
Yang-Mills equations
\begin{equation}
    \Lef D_\mu (\sqrt{-g}F^{\mu\nu})\Rig ^a=0,
\end{equation}
where $D_\mu$ is the total covariant derivative with respect to both
the gauge group and metric connections.
\subsection{Ansatz for the metric and the gauge field}
We shall use the cylindrical coordinates $x^\mu=(t,r,\vf,z)$. The
space-time generated by the static cylindrically symmetric source is
diagonal  and possesses three commuting Killing vectors
$\p_t,\p_\vf,\p_z $. Its explicit form  depends on the chosen gauge,
here we will use  the Kasner gauge  $g_{rr}=-1$:
\begin{equation}
ds^{2}=N^2(r)dt^2-dr^2-L^2(r)d\vf^2-K^2(r)dz^2. \label{eq:MK}
\end{equation}
The corresponding non-zero components of the Ricci tensor are:
\begin{eqnarray}
  R_0^0 &=& \frac{(N'L K)'}{N L K}\;,\qquad
  R_r^r = \frac{N''}{N}+\frac{L''}{L}+\frac{K''}{K}\;,\nonumber\\
  R_{\vf}^{\vf} &=& \frac{(N L' K)'}{N L K}\;,\qquad
  R_z^z = \frac{(N L K')'}{N L K}.
\end{eqnarray}

To construct an appropriate ansatz for the gauge field we have to
ensure that the effect of the field transformation upon the
coordinate translation  along the Killing vector $K$,
\begin{equation}
    A_\mu\rightarrow A_\mu-\epsilon \mathcal{L}_{K}A_\mu,
\end{equation}
can be compensated by a suitable gauge transformation
\cite{Bergmann:1978fi}
\begin{equation}
    A_\mu\rightarrow A_\mu+\epsilon D_\mu W,
\end{equation}
with $W$ being a Lie algebra valued function such that
\begin{equation}
    \mathcal{L}_{K}A_\mu=D_\mu W.\end{equation}
With three Killing vectors we will thus  have for the SU(2) YM
field nine independent functions. Imposing as usual for the static
axially-symmetric configurations a discrete symmetry
$M_{xz}\otimes C$, where the first factor stands for reflection in
the $xz$-plane and the second  is charge conjugation, we reduce
the number of independent components to six. There are different
gauge equivalent ways to parameterize the field potential, we
choose the following gauge:
\begin{eqnarray}
   e A_t &=& \mathrm{T}_r w_{0}^1+\mathrm{T}_z w_{0}^3,\label{eq:At}\\
   e A_r &=& \mathrm{T}_\vf w_{1}^2,\\
   e A_\vf &=& \mathrm{T}_r w_{2}^1+ \mathrm{T}_z w_{2}^3,\\
   e A_z &=& \mathrm{T}_\vf w_{3}^2,\label{eq:Az}
\end{eqnarray}
where the rotated generators $\mathrm{T}_r, \mathrm{T}_\vf$ and
$\mathrm{T}_z$ are used:
\begin{eqnarray}
    \mathrm{T}_n &=& \frac{1}{2 i}\vec{\tau}\vec{e_n}, \qquad \vec{e}_r=(\cos{\nu\vf},\sin{\nu\vf},0),\nonumber\\
    \vec{e}_\vf &=&  (-\sin\nu\vf,\cos\nu\vf,0),\qquad
    \vec{e_z}=(0,0,1),
\end{eqnarray}
$\tau_n$ being the Pauli matrices. The real number $\nu$ must be
integer to avoid multivaluedness of the potential. Two temporal
components in (\ref{eq:At}--\ref{eq:Az}) can be regarded as electric
potentials, while the other  four are the magnetic ones. Thus purely
magnetic static axially symmetric configurations can be described by
four real functions of two variables $w_{i}^a(r,z)$:
\cite{Kleihaus:1996yc,Kleihaus:1997mn}:
\begin{equation}\label{eq:Akk}
   e A_\mu dx^\mu= \mathrm{T}_\vf\Lef w_{1}^2 dr+w_{3}^2 dz\Rig+\Lef \mathrm{T}_r w_{2}^1+ \mathrm{T}_z
    w_{2}^3\Rig d\vf.
\end{equation}
The corresponding gauge fixing condition can be chosen in the form
\begin{equation}\label{eq:gaugec1}
   r \p_r w_{1}^2+\p_z w_{3}^2=0.
\end{equation}

This ansatz has a residual  form-invariance with respect to the
transformation
\begin{equation}\label{eq:gauget1}
U=\exp\Lef \mathrm{T}_\vf \lambda(r,z)\Rig=\cos\frac{\lambda}{2}+ 2
\mathrm{T}_\vf\sin\frac{\lambda}{2},
\end{equation}
under which the functions $(w_{1}^2, w_{3}^2)$ transform as
2-dimensional gauge field
\begin{equation}
    w^{2}_{\alpha}\ra w^{2}_{\alpha}+\p_\alpha \lambda,\qd
    \alpha=1,3,
\end{equation}
while the functions $(w_{2}^1,w_{2}^3-\nu)$  behave as scalar
doublet:
\begin{equation}\label{eq:transform2}
    \left(
\begin{array}{c}
  w_{2}^1 \\
  w_{2}^3-\nu \\
\end{array}
\right)\ra \left(
\begin{array}{cc}
  \cos\lambda & \sin\lambda \\
  -\sin\lambda & \cos\lambda \\
\end{array}
\right)\left(
\begin{array}{c}
  w_{2}^1 \\
  w_{2}^3-\nu \\
\end{array}
\right).
\end{equation}

Now we impose the translational symmetry along the polar axis
assuming that all gauge functions depend only on radial variable:
${w_{i}^a=w_{i}^a(r)}$.  Performing the gauge transformation
\begin{equation}
    U=\exp\Lef \mathrm{T}_\vf \lambda(r)\Rig,\qd \tan\lambda=-\frac{w^{1}_{2}}{w^{3}_{2}-\nu}
\end{equation}
we can exclude the component $w^{1}_{2}$. Furthermore,  the radial
component of the gauge field vanishes by virtue of the YM equation
$w^{2}_{1}+\lambda'=0$. Therefore in the static cylindrically
symmetric magnetic case we are left with only two independent
functions:
\begin{equation}\label{eq:Asym}
    e A_\mu dx^\mu = \mathrm{T}_2 w_{3}^2 dx^3 + \mathrm{T}_3
    w_{2}^3 dx^2.
\end{equation}
It is worth noting that with this parametrization  the equations
of motion  are invariant under discrete transformation
interchanging both the space and color indexes $\vf
\leftrightarrow z$. This symmetry, however,  is broken by the
boundary conditions on the polar axis which is different for
$w_{3}^2$ and $w_{2}^3$. Renaming the  functions $w_{i}^a(r)$ as
in the cosmic string theory, we rewrite the final ansatz in the
form
\begin{equation}\label{eq:ansatz}
e A_\mu dx^\mu=
\mathrm{T}_{\vf}R(r)dz+\mathrm{T}_{z}(P(r)-\nu)d\vf.
\end{equation}
For another way to arrive at this representation see
\cite{Gal'tsov:1998af}.

The corresponding matrix-valued field tensor has the following
non-zero components:
\begin{equation}\label{eq:F}
e F_{r\vf}= \mathrm{T}_z P',\;\; e F_{r z}=
\mathrm{T}_{\vf}R',\;\; e F_{\vf z}=-\mathrm{T}_r PR.
\end{equation}
Subsituting them into the action (\ref{eq:action}) we obtain:
\begin{equation}\label{eq:action2}
    S=\!\int\! d^4x \sqrt{-g}\!
    \left\{\!-\frac{\mathbb{R}}{16 \pi G}-\frac{1}{2e^2}\!\left(\frac{P'^2}{L^2}+\frac{R'^2}{K^2}+
    \frac{P^{2}R^{2}}{L^{2}K^{2}}\right)\!\right\}.
\end{equation}
The stress-energy tensor (\ref{Eq:T1}) for the gauge field is
diagonal and can be expressed in terms of the energy density and
the principal pressures as follows:
\begin{eqnarray}
T_{0}^0 &=& \rho=
\epsilon_\vf+\epsilon_z+\epsilon_w \label{eq:T0},\\
T_{r}^{r} &=& -p_r=-\epsilon_\vf-\epsilon_z+\epsilon_w,\\
T_{\vf}^{\vf} &=& -p_\vf= \epsilon_\vf-\epsilon_z-\epsilon_w,\\
T_{z}^{z} &=& -p_z=-\epsilon_\vf+\epsilon_z-\epsilon_w,
\label{eq:T4}
\end{eqnarray}
 where
\begin{equation}\label{eq:epsilons}
    \epsilon_\vf=\frac{R'^2}{2e^2K^2},\;\;\;
    \epsilon_z=\frac{P'^2}{2e^2L^2},\;\;\;
    \epsilon_w=\frac{P^2 R^2}{2e^2L^2 K^2}.
\end{equation}
Its tracelessness implies \begin{equation}\label{rop}
    \rho=p_r+p_\vf+p_z.
\end{equation}
\subsection{Symmetries of the reduced action}
It is convenient to pass to the dimensionless quantities introducing
the scale factor $l=e^{-1}\sqrt{4\pi G}$ and performing the
rescalings
\begin{eqnarray}
  r &\rightarrow& l^{-1}r,\qquad R \rightarrow l R,\qquad P \rightarrow
  P, \nonumber
  \\
  N &\rightarrow& N,\qquad L \rightarrow l^{-1}L, \qquad K \rightarrow
  K. \label{eq:rescaling}
\end{eqnarray}
Integrating the action (\ref{eq:action2}) over $t, z, \vf$ we
obtain the one-dimensional action
\begin{equation}
    S=\frac{{\rm Vol}(t,z,\vf) }{e^2 l^2}S_1,\quad S_1=\int \mathrm{L} dr
\end{equation}
where ${\rm Vol}(t,z,\vf)$ is the normalization volume, and the
one-dimensional field lagrangian is
\begin{eqnarray}\label{eq:lagrangian1}
    \mathrm{L} &=& \frac{N L K}{2}\Lef\frac{N'L'}{N L}+\frac{N'K'}{N K}+\frac{L'K'}{L K}-\nn\right.\\
    &&\qd\qd \left.-\frac{P'^2}{L^2}-\frac{R'^2}{K^2}-
    \frac{P^{2}R^{2}}{L^2 K^2} \Rig.
\end{eqnarray}
As was argued above, this lagrangian is invariant under a discrete
$\vf\leftrightarrow z$ transformation which now reads as
\extrarowheight=0pt
\begin{equation}\label{eq:gauget3}
    \begin{pmatrix}
      P \\
      L \\
    \end{pmatrix}\Leftrightarrow
    \begin{pmatrix}
      R \\
      K \\
    \end{pmatrix}.
\end{equation}
In terms of the rescaled quantities the stress-energy tensor will
still be given by the Eq. (\ref{eq:T4}) with
\begin{equation}\label{eq:epsilonsr}
    \epsilon_\vf=\frac{R'^2}{2K^2},\quad
    \epsilon_z=\frac{P'^2}{2L^2},\quad
    \epsilon_w=\frac{P^2 R^2}{2L^2 K^2}.
\end{equation}

The dimensional reduction over $t,\;\vf$ and $z$ leads to two
continuous symmetries. To make this more transparent, it is
convenient to use an exponential parametrization  for the metric
and gauge functions:
\begin{eqnarray}
  N &=& e^\nu,\qd L=e^\lambda, \qd K=e^\xi,\nn \\
  R &=& e^\alpha,\qd P=e^\beta.
\end{eqnarray}
In terms of these new variables the lagrangian
(\ref{eq:lagrangian1})
\begin{eqnarray}
  \mathrm{L} &=& \frac{1}{2}e^{\nu+\lambda+\xi}\Lef \nu'\lambda'+\nu'\xi'+\lambda'\xi'-
  e^{2(\beta-\lambda)}\beta'^2- \right.\nn\\
   &&\qd\qd\quad \left. e^{2(\alpha-\xi)}\alpha'^2-
   e^{(\alpha-\xi)+2(\beta-\lambda)}\Rig,
\end{eqnarray}
contains only three independent linear combinations of metric
functions in the exponential.  Choosing them as the new variables:
\begin{eqnarray}
\left\{\begin{array}{rcl}
\tau&=&\nu+\lambda+\xi \\
\sigma&=&-\xi+\alpha \\
\psi&=&-\lambda+\beta \\
\tilde{\alpha}&=&\alpha \\
\tilde{\beta}&=&\beta
\end{array}\right. ,
\end{eqnarray}
one finds that $\tilde{\alpha},\:\tilde{\beta}$ are cyclic.
Therefore the lagrangian is invariant under the infinitesimal
translations of $\tilde{\alpha}$ and $\tilde{\beta}$ which leave
invariant $\tau,\:\sigma$ and $\psi$:
\begin{equation}\label{eq:transform1}
\left\{\begin{array}{rcl}
\delta\alpha&=&0 \\
\delta\beta&=&\ep_1 \\
\delta\psi&=&-\ep_1 \\
\delta\xi&=&0 \\
\delta\lambda&=&\ep_1\end{array}\right., \qd
\left\{\begin{array}{rcl}
\delta\alpha&=&\ep_2 \\
\delta\beta&=&0 \\
\delta\psi&=&-\ep_2 \\
\delta\xi&=&\ep_2 \\
\delta\lambda&=&0\end{array}\right..
\end{equation}
The corresponding Noether charges are
\begin{eqnarray}
  Q_1 &=& \frac{1}{2}e^{\nu+\lambda+\xi}\Lef \nu'-\lambda'- 2e^{2(\beta-\lambda)}\beta'^2\Rig ,\\
  Q_2 &=& \frac{1}{2}e^{\nu+\lambda+\xi}\Lef
  \nu'-\xi'-2e^{2(\alpha-\xi)}\alpha'^2\Rig,
\end{eqnarray}
or, in the previous terms
\begin{eqnarray}
  Q_1 &=& \frac{1}{2}N L K\Lef \frac{N'}{N}-\frac{L'}{L}- 2\frac{PP'}{L^2}\Rig \label{eq:Q1},\\
  Q_2 &=& \frac{1}{2}N L K\Lef \frac{N'}{N}-\frac{K'}{K}-
  2\frac{RR'}{K^2}\Rig. \label{eq:Q2}
\end{eqnarray}
These are conserved on shell together with the total energy
(hamiltonian), which for the static system differs from the
one-dimensional lagrangian by the sign of the potential term:
\begin{eqnarray}\label{eq:hamiltonian1}
    \mathrm{H} &=&\frac{N L K}{2}\Lef\frac{N'L'}{N L}+\frac{N'K'}{N K}+\frac{L'K'}{L K}-\nn\right.\\
    &&\qd\qd \left.-\frac{P'^2}{L^2}-\frac{R'^2}{K^2}+
    \frac{P^{2}R^{2}}{L^2 K^2} \Rig=0.
\end{eqnarray}
\par
Let us discuss some cases studied previously. In some  gauge
 models in curved space-time with cylindrical symmetry, such as
cosmic strings, the stress-energy tensor satisfies the condition
\begin{equation}\label{eq:conictensor}
    T_{0}^{0}=T_{z}^{z},
\end{equation}
implying the full Poincare symmetry of the $(t,z)$-plane. In this
case  $g_{tt}= g_{zz}$ and  the metric is described by only two
functions. From the Eq. (\ref{eq:T4}) it is clear that this
corresponds to $\epsilon_z=0$, i.e. $P=\nu$ (to ensure the
regularity of the axis ), in which case we  find the Melvin
solution \cite{Melvin:1963qx}, which we will call here
``$z$-string''. In the general case, in view of the $z-\vf$
symmetry mentioned above, we deal with the non-linear
superposition of the $z$-string and the $\vf$-string. These are
interacting via the non-linear terms in the Yang-Mills equations.
\par
It is easy to see that a single $\vf$-string is singular on the
axis $r=0$. Moreover, within the Einstein-Maxwell theory a
configuration which contains both strings inherits the
$\vf$-string singularity. However  the self-interaction in the
non-Abelian theory   makes possible the superposition of both
strings regular on the polar axis. As we will see, in such
configurations the contribution of the $\vf$-string is suppressed
near the polar axis, this ensures the regularity of the full
solution.
\section{Equations of motion and conditions on the polar axis}
In view of the tracelessness of the stress-energy tensor of the
gauge field  we have three independent Einstein equations which read
(\ref{eq:Einsteq1})
\begin{eqnarray}
\frac{(N'L K)'}{N L K} &=& \frac{P'^2}{L^2}+
\frac{R'^2}{K^2}+\frac{P^{2}R^{2}}{L^{2}K^{2}} \label{eq:Ncurved},\\
\frac{(N L' K)'}{N L K} &=& -\frac{P'^2}{L^2}+
\frac{R'^2}{K^2}-\frac{P^{2}R^{2}}{L^{2}K^{2}}, \\
\frac{(N L K')'}{N L K} &=& \frac{P'^2}{L^2}-
\frac{R'^2}{K^2}-\frac{P^{2}R^{2}}{L^{2} K^{2}}.\label{eq:Kcurved}
\end{eqnarray}
In the Kasner gauge (\ref{eq:MK})   the ($rr$) Einstein equation is
a first order equation which is the equivalent to the hamiltonian
constraint (\ref{eq:hamiltonian1}).

The Yang-Mills equations have a symmetric form
\begin{eqnarray}
\left(R'\frac{N L}{K}\right)' &=& \frac{N}{L K}R P^2 \label{eq:Rcurved},\\
\left(P'\frac{N K}{L}\right)' &=& \frac{N}{L K}P R^2
\label{eq:Pcurved}.
\end{eqnarray}
As a whole we have the system of five non-linear second-order
differential equations (\ref{eq:Ncurved}--\ref{eq:Pcurved}), which
must be completed by the set of boundary conditions. In this
paper, we will be interested in the globally regular
asymptotically Ricci-flat solutions. So  we have to formulate the
regularity conditions on the symmetry axis and to analyze the
asymptotic behavior compatible with the ansatze for the metric and
the YM field.  The regularity conditions on the polar axis will be
discussed separately for the flat-space problem and within the
full self-gravitating treatment.
\subsection{Expansion near the polar axis in flat space}
In the flat space-time $N=K=1,\:L=r$, and the equations of motion
(\ref{eq:Rcurved}), (\ref{eq:Pcurved}) read
\begin{eqnarray}
r\left(r R'\right)'=RP^2 \label{eq:Req},\\
r\left(\frac{P'}{r}\right)'=R^2 P \label{eq:Peq}.
\end{eqnarray}
For the system of two second-order differential equations we have
to impose  four boundary conditions on the polar axis. Requiring
the solutions to be regular on the polar axis,  we find the
conditions $PR=0$ at $r=0$:
\begin{equation}\label{eq:RPcond}
P(0)=k,\qd R(0)=0,
\end{equation}
where $k$ is a real parameter. We will  consider $k\neq 0$, since
for $k=0$ and $R(0)=\rm{const}$ the solution is trivial
$R(0)=P(0)=0$.

Note that in the Abelian case the right hand side of the equation
(\ref{eq:Rcurved}) is  zero. Integrating this equation one finds a
singularity $R'(r)=R'(0)/r$ which can be avoided only by excluding
the $\vf-$component of the gauge field from the ansatz. In the
non-Abelian case this singularity is avoided by imposing the above
conditions.

To proceed further, we look for the power series solutions for $R$
and $P$ near the polar axis. The first coefficient in the right
hand side of the equation for $R$ is proportional to $k^2$. Thus
in the left hand side of the equation the leading term must be
proportional to $r^{k}$. One can find that the other terms in the
expansions for $R$ and $P$ are proportional to $r^{2^nk+m}$, where
$n$ and $m$ are integers. In the lowest order one obtains the
following expansion:
\begin{eqnarray}
  R &=& \frac{q}{[k]!}r^k+\frac{q p k}{4(k+1)}r^{2+k}+O(r^{2+2k}), \label{eq:expRk}\\
  P &=&
  k+\frac{pr^2}{2}+\frac{q^2}{4(k+1)}r^{2+2k}+O(r^{3+2k}), \label{eq:expPk}
\end{eqnarray}
where $[k]$ denotes the integer part of  $k$. The free parameters
$q$ and $p$ represent the boundary conditions to the derivatives
on the axis $r=0$. Together with (\ref{eq:RPcond}) one has four
boundary conditions which are sufficient to formulate the boundary
problem.

For an integer $k$ the parameters $q$ and $p$ are equal to the
higher order derivatives of $R$ and $P$ as follows
\begin{equation}\label{eq:RPcond2}
    p=P''(0),\qd q=R^{(k)}(0),
\end{equation}
and the expansions contain only integer powers of $r$, therefore
the solutions for $R$ and $P$ are analytic functions. In this case
we can obtain the higher order terms in series of expansions which
are presented in Tab.~{\ref{tb:flat}}. These have the following
structure:
\begin{eqnarray}
  R &=& \sum\limits_{j=0}^{\infty}R_{k+2j}\frac{r^{k+2j}}{(k+2j)!}, \label{eq:expR}\\
  P &=&
  k+p\frac{r^2}{2}+\sum\limits_{j=0}^{\infty}P_{2(k+j+1)}\frac{r^{2(k+j+1)}}{{2(k+j+1)}!},\label{eq:expP}
\end{eqnarray}
where the coefficients  of any order can be expressed through the
free parameters $p, q$ as follows
\begin{eqnarray}
  R_{k+2j} &=& \sum\limits_{i=0}^{\left[\frac{j}{k+1}\right]}C^{(k)}_{ij}q^{2i+1}p^{j-i(k+1)},\label{eq:Rn} \\
  P_{2(k+j+1)} &=&
  \sum\limits_{i=0}^{\left[\frac{j}{k+1}\right]}D^{(k)}_{ij}q^{2(i+1)}p^{j-i(k+1)}.\label{eq:Pn}
\end{eqnarray}

Thus we have the three-parameter family of solutions to the system
(\ref{eq:Req},~\ref{eq:Peq}) specified by the set $\{q,~p,~k\}$.
But as will be shown below one can fix e.g. $q=1$ and consider
only the two-parameter family $\{p,~k\}$, while other solutions
with $q\neq 1$ can be obtained by transformations of the family
$q=1$.

It is easy to see that if $\{R(r),~P(r)\}$ is a solution to the
system (\ref{eq:Req},~\ref{eq:Peq}) then $\{lR(lr),~P(lr)\}$ will
be also a solution, where $l$ is some constant. The question
arises whether this rescaling affect parameters $\{q,~p,~k\}$.
Consider the solution $\{R(q_0,p_0,k_0;r),~P(q_0,p_0,k_0;r)\}$ and
make the scale transformation
$\{lR(q_0,p_0,k_0;lr),~P(q_0,p_0,k_0;lr)\}$. The latter solution
must coincide with some solution
$\{R(q_l,p_l,k_l;r),~P(q_l,p_l,k_l;r)\}$. In the leading orders of
expansion one has
\begin{eqnarray}
  q_lr^k &=& lq_0r^kl^k, \\
  k_l+p_lr^2/2 &=& k_0+p_0r^2l^2/2.
\end{eqnarray}
This implies that $k_l=k_0=k$, $p_l=pl^{-2}$ and
$l=(q_l/q_0)^{\frac{1}{k+1}}$. Extracting $l$ one obtains
\begin{equation}\label{eq:pq}
    p=p_0(q/q_{0})^{\frac{2}{k+1}}.
\end{equation}
Thus all solutions belonging to the curve (\ref{eq:pq}) in the
plane $k=\rm{const}$ of the parameter space can be obtained from
the solution $\{q_0=1,~p_0,~k\}$ only by the scale transformation
\begin{eqnarray}
  R(q,p,k;r) &=& q^\frac{1}{k+1}
  R\left(1,pq^{-\frac{2}{k+1}},k;q^\frac{1}{k+1}r\right), \\
  P(q,p,k;r) &=& P\left(1,pq^{-\frac{2}{k+1}},k;
  q^\frac{1}{k+1}r\right).
\end{eqnarray}
In the following we will restrict to the case $q=1$.

\subsection{Curved space}
Near the polar axis the metric component $g_{\vf\vf}=L^2(r)$ must
vanish. To ensure the finiteness of the curvature the regularity
conditions on the metric functions and their derivatives must be
imposed. To find them we consider the Kretchmann scalar,
\begin{eqnarray}\label{eq:krechman}
    R_{\mu\nu\lambda\tau}R^{\mu\nu\lambda\tau} &=&\left(\frac{N'L'}{N L}\right)^2+
    \left(\frac{N'K'}{N K}\right)^2+\left(\frac{L'K'}{L
    K}\right)^2+
    \nn\\
    &&\left(\frac{N''}{N}\right)^2+\left(\frac{L''}{L}\right)^2+
    \left(\frac{K''}{K}\right)^2,
\end{eqnarray}
which must be finite  on the polar axis. Since $L$ vanishes there,
one has to require $L''(0)=0$ as well. This implies that $L'(0) $
must be non-zero, otherwise $g_{\vf\vf}$ will be identically zero.
Furthermore, vanishing of  denominators  in the upper string of
(\ref{eq:krechman}) must be compensated by vanishing of
numerators, which implies $N'(0)=K'(0)=0$. Finally, $N$ and $K$
cannot vanish  on the axis or they will be equal to zero
everywhere. Therefore the locally flat metric starting with
\begin{eqnarray}
    N(0)&=&K(0)=1,\qd L(0)=0, \nn\\
    N'(0)&=&K'(0)=0,\qd L'(0)=1.\label{eq:NLKcond}
\end{eqnarray}
is the only  one (up to rescaling) which provide finiteness of the
expression (\ref{eq:krechman}) one the polar axis. Note that from
the azimuthal Einstein equation  one can show that the condition
\begin{equation}\label{eq:Lcond}
    L''(r=0)=0
\end{equation}
is satisfied if the azimuthal pressure is finite near the polar
axis.
\par
Now we use the same strategy to find the appropriate condition on
the gauge field (\ref{eq:ansatz}) constructing the power series
expansions near $r=0$ for all variables satisfying the metric
conditions (\ref{eq:NLKcond})  directly from the system
(\ref{eq:Ncurved}--\ref{eq:Pcurved}). One can expect  similar
results for the field functions as in the  flat case with some
amendments due to gravity. This is indeed the case as can be seen
from  Tab.~\ref{tb:curved}. But the expansions of the metric
functions for non-integer $k$ read
\begin{eqnarray}
  N &\sim& 1+C_N r^{2k}+O(r^{2k+2}), \\
  K &\sim& 1+C_K r^{2k}+O(r^{2k+2}).
\end{eqnarray}
Obviously the terms in the Kretchmann scalar (\ref{eq:krechman})
which are proportional to the second derivatives of $N$ and $K$
$\sim r^{2k-2}$ vanish in vicinity of the axis only when $k\geq
1$. The energy density and pressure, which are proportional to
$r^{2k-1}$ in the vicinity of the axis, vanish in this case.
Therefore the condition (\ref{eq:Lcond}) is also satisfied.

Note that the gravity coupled solution does not contain any new
free parameters. The full set of ten boundary conditions consists
of the previously obtained two conditions for the field functions
(\ref{eq:RPcond}) and the  six conditions on the metric
(\ref{eq:NLKcond}), with three parameters $\{p,\:q, \: k\}$
remaining free (recall that to ensure analyticity of the solution
near the polar axis one has to take integer $k$).
\par
Unlike the flat case, now the series expansions contain some
additional terms. But all such terms include only higher powers of
parameters $p$ and $q$ apart from the main terms from
(\ref{eq:expRk}), (\ref{eq:expPk}). For this reason in the weak
field case $p\rightarrow 0,\: q\rightarrow 0$ the curved space
expansions are dominated by the flat space ones (\ref{eq:expRk}),
(\ref{eq:expPk}), i.e.  the solutions satisfying (\ref{eq:pq})
will differ only by the scale factors. This scale invariance does
not hold in general, however.

Now we would like to clarify the physical meaning of the parameter
$p$. Using the Tab.~\ref{tb:curved} one finds
\begin{equation}
    p=\lim\limits_{r\rightarrow 0} \frac{P'(r)}{L(r)}.
\end{equation}
Comparing this expression with the  component $F_{r\vf}$ of the
field tensor (\ref{eq:F}), one can see that this quantity is
nothing else than the corresponding tetrad component with the
color index $z$. It is therefore the limiting value of the
physical  magnetic field component along the symmetry axis
\begin{equation}\label{eq:B}
    B_z(r)=-\frac{P'(r)}{eL(r)}.
\end{equation}
After rescaling, we obtain the dimensionless parameter
\begin{equation}\label{eq:B0}
\mathbb{B}=4\pi G e^{-1}B(0),
\end{equation}
therefore the parameter $p$ is equal up to sign to the
(dimensionless) value of $z$-component of the magnetic field on
the polar axis:
\begin{equation}\label{pB}
    p=-\mathbb{B}.
\end{equation}

\subsection{Sum rules}
To relate the  expansion on the symmetry axis to the asymptotic
form of the solution on can derive the sum rules following the
approach of the Ref. \cite{Colding:1997bw}. Multiplying the
Einstein equations (\ref{eq:Ncurved02}--\ref{eq:Kcurved02}) by
\begin{equation}\label{det}
    \sqrt{-g}=NLK,
\end{equation}
we integrate them from $r=0$ to $r$ using the expressions
(\ref{eq:T4}, \ref{eq:epsilonsr}) for the stress-energy tensor.
This gives
\begin{eqnarray}
  N' L K &=& M, \label{eq:Ncurved02}\\
  N L' K &=& W +1, \\
  N L K' &=& X, \label{eq:Kcurved02}
\end{eqnarray}
where three new functions are introduced to the integrals
\begin{eqnarray}
 M &=& \int\limits_{0}^{r}dr \label{eq:M1}
 \sqrt{-g}(\rho+p_r+p_\varphi+p_z), \\
 W &=& -\int\limits_{0}^{r}dr
 \sqrt{-g}(\rho-p_r+p_\varphi-p_z),\\
 X &=& -\int\limits_{0}^{r}dr
\sqrt{-g}(\rho-p_r-p_\varphi+p_z).\label{eq:X1}
\end{eqnarray}
The function $M(r)$ is the Tolman mass of the gravitating YM field
within  the cylinder of the radius $r$. Two other quantities are
related to integral pressures in the longitudinal and azimuthal
directions, their more precise meaning will be clear later on. The
sum rules (\ref{eq:Ncurved02}-\ref{eq:Kcurved02}) express the
metric functions through the integral characteristics of the
matter field.

Two other some rules can be obtained by integrating the YM
equations. Multiplying (\ref{eq:Rcurved}, \ref{eq:Pcurved}) on $R$
and $P$ respectively, we can rewrite the YM equations as follows
\begin{eqnarray}
\left(\frac{N L}{K}R R'\right)' &=& N L K
\left(\frac{R'^2}{K^2}+\frac{P^2 R^2}{L^2 K^2}\right),\label{eq:Rcurved2}\\
\left(\frac{N K}{L}P P'\right)' &=& N L K
\left(\frac{P'^2}{L^2}+\frac{P^2 R^2}{L^2 K^2}\right)
\label{eq:Pcurved2}.
\end{eqnarray}
Then we integrate in the same way as the Einstein equations
obtaining the sum rules:
\begin{eqnarray}
  \frac{N L}{K} \frac{(R^2)'}{2} &=& \int\limits_{0}^{r}
  dr \sqrt{-g}\left(\frac{R'^2}{K^2}+\frac{P^2 R^2}{L^2 K^2}\right)=\nn\\
  &&\qd\qd\qd\qd=
  \frac{M-X}{2} \label{eq:Rcurved3},\\
  \frac{N K}{L} \frac{(P^2)'}{2} &=& \int\limits_{0}^{r}
  dr \sqrt{-g}\left(\frac{P'^2}{L^2}+\frac{P^2 R^2}{L^2
  K^2}\right)+\nonumber\\ &\lim\limits_{r\rightarrow 0}&\left(N K
  P\frac{P'}{L}\right)=
  \frac{M-W}{2}- k\mathbb{B} \label{eq:Pcurved3}\;.
\end{eqnarray}
From the three integral functions $M,\, W,\, X$ only two are
independent: the constraint equation (\ref{eq:hamiltonian1})
implying
\begin{equation}\label{eq:MWXcons}
    M X+M(W+1)+X(W+1)=0.
\end{equation}
\section{Asymptotic behavior}
Methods of investigations of the asymptotical geometry generated
by cylindrically symmetric field configurations were developed
within the model of Abelian cosmic strings \cite{Garfinkle:1985hr,
Frolov:1989er, Verbin:1998tc, Christensen:1999wb}, for a more
general model-independent approach see \cite{Colding:1997bw}. Here
we shall use the same technique with some modifications
appropriate to the YM field dynamics.

\subsection{Kasner solution}
It is well known that the general vacuum solution of Einstein
equations for the  ansatz (\ref{eq:MK}) is given by the Kasner
metric
\begin{equation}\label{eq:metrika_kasner}
ds^2=r^{2a}dt^2-dr^2-\gamma^2 r ^{2b}d{\vf}^2-r^{2c}dz^2,
\end{equation}
where constants $a,\:b$ and $c$ satisfy  the conditions
\begin{equation}\label{eq:kascond}
a+b+c=a^2+b^2+c^2=1.
\end{equation}
Consider the cylindrically symmetric configurations of the YM
field for which the stress-energy tensor vanishes at $r\ra\infty$.
Then the metric in the asymptotical region will approach
(\ref{eq:metrika_kasner}), where parameters $a,\:b$ and $c$ must
be specified in terms of parameters of the field configuration.
For this purpose we define the asymptotic values of the integral
functions $M,\,W,\,X$ assuming them to be finite
\begin{eqnarray}\label{Minfty}
  M_\infty &=& \lim_{r\ra\infty}M(r), \nonumber\\
  W_\infty &=& \lim_{r\ra\infty}W(r), \\
  X_\infty &=& \lim_{r\ra\infty}X(r),\nonumber
\end{eqnarray}
and denote $\Sigma_\infty=M_\infty+W_\infty+X_\infty+1$. These
three parameters satisfy a constraint (\ref{eq:MWXcons}):
\begin{equation}\label{eq:kascond2}
    M_\infty X_\infty+M_\infty(W_\infty+1)+X_\infty(W_\infty+1)=0.
\end{equation}
Together with $\mathbb{B}$, we thus have four parameters to be
related to the Kasner parameters. From the Eqs.
(\ref{eq:Ncurved02}--\ref{eq:Kcurved02}) it follows
\begin{eqnarray}
  N' L K &\ra& M_\infty \label{eq:Ninfty},\label{eq:N1}\\
  N L' K &\ra& W_\infty +1, \\
  N L K' &\ra& X_\infty. \label{eq:Kinfty}\label{eq:K1}
\end{eqnarray}
It is easy to find the solution to this equations assuming
$\Sigma_\infty \neq 0$:
\begin{eqnarray}
  N &=& \omega r^{a}\;,\qquad a=M_\infty/\Sigma_\infty \label{eq:Ncurved2},\\
  L &=& \gamma r^{b} \;,\qquad b=(W_\infty+1)/\Sigma_\infty \label{eq:Lcurved2},\\
  K &=& (\omega \gamma)^{-1}r^{c}\;,\qquad c=X_\infty/\Sigma_\infty \label{eq:Kcurved2},\\
  \sqrt{-g} &=& N L K =\Sigma_\infty r \label{eq:NLKcurved2},
\end{eqnarray}
where $\omega, \gamma$ are some constants. The first Kasner
condition (\ref{eq:kascond}) is obviously satisfied , while the
second is equivalent to (\ref{eq:kascond2}).

So we are left with only two independent parameters describing the
asymptotically vacuum metric. The condition (\ref{eq:kascond2})
means the fulfilment of the radial Einstein equation. From
(\ref{eq:NLKcurved2}) it is clear that the condition
$\Sigma_\infty \neq 0$ is the requirement of the non-degeneracy of
the metric at infinity. On the polar axis $r=0$ the boundary
condition for $\sqrt{-g}$ read that its derivative is positive.
Since $\sqrt{-g}$ must be non-vanishing at any point except $r=0$,
one has to require $\Sigma_\infty>0$.
\subsection{Existence of asymptotically vacuum solutions}
Now let us turn to the dynamic of the gauge field. The assumption
of the finiteness of $M_\infty,\:W_\infty$ and $X_\infty$ imposes
some conditions on the gauge field asymptotics. Actually, the
energy density must decrease fast enough for $M_\infty$ to be
finite. From Eqs.(\ref{eq:M1}---\ref{eq:M1}) it is clear that the
absolute values of $W_\infty$ and $X_\infty$ are smaller than
$M_\infty$. The total energy density is the sum of three
components (\ref{eq:T0}), which asymptotically can be written as
\begin{equation}\label{eq:epsilons2}
    \epsilon_\vf\sim R'^2 r^{-2c},\;\;
    \epsilon_z\sim P'^2 r^{-2b},\;\;
    \epsilon_w\sim P^2 R^2 r^{2a-2}.
\end{equation}
All these quantities must fall faster than $r^{-2}$ to ensure
finiteness of $M_\infty$. Notice that
$0\!<\!a=\!M_\infty/\Sigma_\infty\leq \!1$ since
$\Sigma_\infty>0$, and $M_\infty>0$ because it is proportional to
the energy of the configuration per unit length. Therefore
$2a-2>-2$ and $RP=o(1)$ as $r\ra\infty$.

Now we can obtain further information  invoking the sum rules
(\ref{eq:Rcurved3},~\ref{eq:Pcurved3}) in combination with the
metric asymptotics (\ref{eq:Ncurved2}--\ref{eq:Kcurved2}):
\begin{eqnarray}
  (R^2)' &=& \frac{(M_\infty-X_\infty)K^2}{\Sigma_\infty r} \label{eq:Rcurved4},\\
  (P^2)' &=&
  \frac{(M_\infty-W_\infty-\mathbb{B})L^2}{\Sigma_\infty r}.\label{eq:Pcurved4},
\end{eqnarray}
Integrating these equations in the leading order with account for
\ref{eq:Ncurved2}--\ref{eq:Kcurved2}) we obtain
\begin{eqnarray}
  R^2 &=& C_{1}^{(R)}+C_{2}^{(R)} r^{2 c}, \label{eq:Rcurved5}\\
  P^2 &=& C_{1}^{(P)}+C_{2}^{(P)} r^{2 b}, \label{eq:Pcurved5}
\end{eqnarray}
with four constant parameters, which can be determined using
(\ref{eq:epsilons2}). The integrand in (\ref{eq:Rcurved3}) is
positively definite, so the integral is non-zero. Since
$C_{2}^{(R)}$ is proportional to this integral, it must be
non-zero too. To satisfy the finiteness condition for $M_\infty$
with asymptotic behavior of the first exponents in
Eqs.(\ref{eq:epsilons2}) the quantity $2(2c-1)-2c=2c-2$ has to be
smaller than $-2$, so $c$ must be negative. Since among the set of
parameters $a,\:b$ and $c$ satisfying the Kasner conditions there
can not be two negative quantities, $b$ has to be positive and the
corresponding exponent in Eqs.(\ref{eq:epsilons2}) is equal to
$2b-2>-2$. Thus to ensure the finiteness condition for $M_\infty$
we must take $C_{2}^{(P)}=0$, or, in other terms,
\begin{equation}\label{eq:Bcond2}
\int\limits_{0}^{\infty} dr N L K\left(\frac{P'^2} {L^2}+\frac{P^2
R^2}{L^2 K^2}\right)-k\mathbb{B}=0,
\end{equation}
what is equivalent to one more constraint on the parameters:
\begin{equation}\label{eq:Bcond20}
    M_\infty-W_\infty=2k\mathbb{B}.
\end{equation}
Note that a similar relation (up to some factor) was obtained
within the model of Abelian cosmic strings
\cite{Christensen:1999wb} as the condition of existence of the
Melvin-like solutions.
\par
To get further information about asymptotics of the gauge
functions we apply  the same integration procedure as described in
section III  to the non-transformed equation of motion
(\ref{eq:Rcurved}). This gives
\begin{equation}\label{eq:Rcurved6}
    R=C_1+C_2 r^{2c}
\end{equation}
with two more constants. Since $c<0$, for the square roots of
(\ref{eq:Rcurved5}) one has two possibilities:
\begin{eqnarray}
  R &=& \sqrt{C_{1}^{(R)}} r^{c},\qd C_{1}^{(R)}=0, \\
  R &=& \sqrt{C_{1}^{(R)}}+\frac{C_{2}^{(R)}}{2\sqrt{C_{1}^{(R)}}} r^{2c},\qd C_{1}^{(R)}\neq 0.
\end{eqnarray}
Comparing with (\ref{eq:Rcurved6}) we see that the second case is
realized. But this implies that $\qd C_{1}^{(P)}=0$, and $P\ra 0$
at infinity faster than power-law, while $R$ approaches a constant
as a power with the exponent $2c$. Finally one  finds the
following explicit expression for $C_{2}^{(R)}$:
\begin{equation}
    C_{2}^{(R)}=\frac{M_\infty-X_\infty}{4C_1X_\infty\omega^2\gamma^2}.
\end{equation}
\par
Now consider the flat space case.  We are looking for
asymptotically vacuum configurations, which implies vanishing
energy density at infinity:
\begin{equation}\label{eq:RPcond7}
    rR'^2+\frac{P'^2}{r}+\frac{R^2P^2}{r} \ra 0.
\end{equation}
Integrating the equations of motion (\ref{eq:Req}), (\ref{eq:Peq})
with zero right hand sides, one obtains the asymptotical
expressions for field functions
\begin{eqnarray}
  R &=& C_1\ln r+C_1, \label{eq:Ra0}\\
  P &=& C_3 r^2+C_4,
\end{eqnarray}
which satisfy (\ref{eq:RPcond7}) if $C_3=0$. Also there must be
$C_4=0$ otherwise the right hand side in the equation for $P$ will
be non-zero. Thus the conditions for the field components at
infinity read
\begin{equation}\label{eq:RPcondflat}
    R\sim \ln r,\quad P\ra {\rm 0}.
\end{equation}
These are the necessary conditions of vanishing of the energy
density. For more detailed expansion see sec.V.

Performing the same steps with the equation for $P$, one obtains
the following relation:
\begin{equation}\label{eq:Bcondflat0}
 \lim\limits_{r\rightarrow \infty}   \frac{(P^2)'}{2r}= \int\limits_{0}^{\infty}dr r\left(\frac{P'^2}{r^2}+
    \frac{R^2 P^2}{r^2}\right)-\lim\limits_{r\rightarrow 0}
  P\frac{P'}{r}.
\end{equation}
Therefore the condition (\ref{eq:RPcondflat}) is satisfied if the
integral exists and the whole expression is equal to zero. This is
the necessary condition of the existence of asymptotically vacuum
solutions:
\begin{equation}\label{eq:Bcondflat}
    \int\limits_{0}^{\infty}dr r\left(\frac{P'^2}{r^2}+
    \frac{R^2 P^2}{r^2}\right)-k\mathbb{B}=0.
\end{equation}
In the flat case all values $M(r)$, $W(r)$ and $X(r)$ diverges as
$\ln{r}$ at infinity because of the term $R'^2/r\sim r^{-1}$ in
the integrand. But the difference $(M-W)$ remains finite and it is
equal to the integral in  the left hand side of
Eq.(\ref{eq:Bcondflat}). So the form of the bifurcation condition
(\ref{eq:Bcond20}) remains the same though the separate values of
$M_\infty$ and $W_\infty$ are not well defined.
\subsection{Kasner parameters}
So far we have obtained the asymptotic behavior  of the gauge and
metric functions consistent  with the equations of motion. The
solution for the metric is determined by three parameters $a,\:b$
and $c$ satisfying two constraints. Now recall that there are also
two conserved  Noether charges (\ref{eq:Q1}), (\ref{eq:Q2}). This
enables us to expose the asymptotic parameters of the  solution
through the data on the polar axis.
\par
Equating the Noether charges $Q_i|_{r=0}=Q_i|_{r\ra\infty}$ one
obtains further two constraints:
\begin{eqnarray}
2k\mathbb{B}-1 &=& \Sigma_\infty(a-b),\\
  0 &=&
  \Sigma_\infty\left(a-c-\frac{M_\infty-X_\infty}{\Sigma_\infty}\right).
\end{eqnarray}
These relations can be  used to express $b$ and $c$ in terms of
two new parameters:
\begin{eqnarray}
  b &=& a+\beta,\qd \beta=\frac{1-2k\mathbb{B}}{\Sigma_\infty},\label{eq:beta}\\
  c &=& a-\sigma,\qd \sigma=\frac{M_\infty-X_\infty}{\Sigma_\infty}.
\end{eqnarray}
Now, from the first Kasner condition one has
\begin{equation}
    a=\frac{1}{3}(1+\sigma-\beta).
\end{equation}
Finally, the quadratic Kasner condition allows us to express all
parameters in terms of the only parameter $\beta$:
\begin{eqnarray}
  a &=& \frac{1}{3}(1-3(\beta/2)+\sqrt{1-3(\beta/2)^2}), \label{eq:a}\\
  b &=& \frac{1}{3}(1+3(\beta/2)+\sqrt{1-3(\beta/2)^2}), \\
  c &=& \frac{1}{3}(1-
2\sqrt{1-3(\beta/2)^2}). \label{eq:c}
\end{eqnarray}
In the previous subsection we have showed that $c$ must be
negative, this determines the sign of square root, an alternative
choice of a sign would give $c\geq 1/3$. We will show later that
the expression under the square root is always positive.
\par
Thus the value $\mathbb{B}$ of the magnetic field value on the
symmetry axis along with $\Sigma_{infty}$ fully specify the
solution in the asymptotic region.

\subsection{Asymptotic YM vacua}
As was shown above the real number $k$ specifies the family of
solutions for $R$ and $P$ i.e. the radial part of the ansatz.
Since $R\sim r^k$ in the vicinity of the axis, the azimuthal
component of the magnetic field $B_\vf \sim R'\sim r^{k-1}$
diverges for non-integer $k$ on the interval $0<k<1$ and so does
the Kretchmann scalar (\ref{eq:krechman}). To ensure analyticity
of the solution near the polar axis $k$ should be chosen integer.
Thus $k$ can be regarded as a "radial" number of the
configuration. If $k=0$ there are no regular solutions except the
trivial one, since the condition (\ref{eq:Bcond20}) implies
vanishing of the energy density.

The parameter $\nu$ does not enter the equations of motion. This
parameter must be integer to avoid the multivaluedness of the
solution. Actually it (and only it) specifies the azimuthal part
of the ansatz, while $k$ (and only $k$) specifies the radial part.
The full analytic solution comes as a product of both parts and
thus is characterized by two integers. The natural question arises
whether the  integers $\nu$ and $k$ can be given any topological
interpretation.

Using the asymptotic expansion for  the gauge potential $A_\mu$
(recall that for large  $r$ the function $P$ vanishes while $R$
tends to a constant, $C$), we find asymptotically the potential
one-form as
\begin{equation}\label{eq:A_asym1}
    A|_{\infty}=C\mathrm{T}_\vf dz-\nu\mathrm{T}_z d\vf,
\end{equation}
and the field two-form  $F=dA+A\wedge A$ vanishes. This implies
that $A$ in (\ref{eq:A_asym1}) describes a vacuum state and  can
be  expressed through the matrix-valued gauge function:
\begin{equation}\label{eq:AU}
    A|_{\infty} =U d U^{-1}.
\end{equation}
This equation can be solved as follows:
\begin{equation}\label{eq:U_asym1}
    U=\frac1{\sqrt{2}}\left(\cos f(1-2\mathrm{T}_r)-
    2\sin f(\mathrm{T}_\vf+\mathrm{T}_z)\right),
\end{equation}
where \begin{equation}\label{eq:f}f=(Cz-\nu\vf)/2.\end{equation}

Now consider the potential on the polar axis. Here it is also a
pure gauge:
\begin{equation}\label{eq:A_asym0}
    A|_{\rm axis}=(k-\nu)\mathrm{T}_z d\vf=\tilde{U}d \tilde{U}^{-1}
\end{equation}
with
\begin{equation}\label{eq:U_asym01}
    \tilde{U}=\cos \tilde{f}-2\sin \tilde{f}
    \mathrm{T_z},\qd \tilde{f}=(k-\nu)\vf/2.
\end{equation}
These two vacua are related by the gauge transformation
(\ref{eq:U_asym1}) with $f=(Cz-k\vf)/2$. The winding number of
this transformation is  zero and the transformation is ``small''.
This reflects the fact that the mapping of the asymptotic cylinder
$(r\ra\infty,~z)$ into the gauge group SU(2)$\sim S^3$ is always
topologically trivial. Thus we can conclude that the integers
$\nu$ and $k$ do not have any topological meaning. Note also that
the divergence of the Chern--Simons current
\begin{equation}
    \nabla_\mu K^\mu=\frac{e^2}{16\pi^2}{\rm tr} F_{\mu\nu}\tilde{F}^{\mu\nu}
\end{equation}
is   zero because of the absence of electric part in the ansatz.

It is instructive to compare the situation with that in the case
of the EYM sphalerons of Bartnik-McKinnon's type. Sphalerons are
the finite energy saddle points in the configuration space of the
system, whose existence can be revealed by the constructing the
non-contractible loops of configurations \cite{Manton} which
alternatively can be seen  as connecting topologically distinct
vacua. For the spherically symmetric four-dimensional EYM
configurations this was done explicitly in \cite{GalVol}.
Configurations of finite energy along the loop form a manifold
homeomorphic to $S^3$, and the existence of non-contractible loops
is a consequence of non-triviality of the mapping of this $S^3$ to
the gauge group $SU(2)\sim S^3$. In the cylindrical case one
considers the loop of configurations of finite energy per unit
length which form the manifold equivalent to $S^2$. Thus instead
of the mapping $S^3\to S^3$ one deals with the mapping $S^2\to
S^3$ which is topologically trivial. Therefore in the cylindrical
case the solutions do not admit sphaleronic interpretation.

\section{Existence of global solutions and numerical integration}
In the previous sections we have shown that global solutions
asymptotically approaching  vacuum configurations, if exist,
should be fully specified by the  boundary conditions on the polar
axis. Up to rescaling,  the only relevant quantity is the
parameter $p=-\mathbb{B}$ which is the second derivative of $P$ on
the axis (see Sec. III, we set the scale taking $q=1$). It turns
out that the family of solutions specified by different $p$
exhibits the bifurcation structure similar to that observed in the
Bartnik-McKinnon case, so the existence of regular solutions can
be proved by similar methods. Here we sketch a proof and present
the results of numerical integration which strongly supports this
analysis.

\subsection{ Flat case }
To  better understand  the dynamics of the gauge field  we start
with the flat-space problem. Our goal is to investigate the
bifurcating behavior of the system (\ref{eq:Req},~\ref{eq:Peq}).
Unfortunately the dynamical systems technique does not provide
enough information about solutions. There are three special points
of the system:  two at the the origin $(r=0,P=0)$ and $(r=0,R=0)$
and one at infinity $(1/r=0,P=0)$. The corresponding linearized
systems are degenerate and   phase portraits in the vicinity of
the special points are trivial. It turns out that the values of
parameters $p$ and $q$ are non-small on the bifurcation curve, so
we can not use the asymptotic methods for the analysis. Our
reasoning will based on the fact that the  equations
(\ref{eq:Req},~\ref{eq:Peq}) are linear with respect to the
variable entering with its derivative, if the other variable is
considered as a given function.
\par
First, let us  investigate   properties of the solutions of the
Eq.(\ref{eq:Req}) with an arbitrary regular function $P(r)$. In
what follows we assume $P(0)=k>0$.
\begin{lem}
The solution  for $R(r)$ with the boundary conditions $R(0)=0$,
$R'(0)=q>0$ is monotonously increasing function for $r>0$.
\end{lem}
\begin{proof}
Since $R'(0)>0$, it remains positive in some region $0<r<a$.
Assuming that it vanishes at  $r=a$, we integrate the
Eq.(\ref{eq:Req}) on the segment $[0,a]$:
\begin{equation}\label{eq:Ra12}
    R'(a)=\frac{1}{a}\int\limits_{0}^{a}\frac{RP^2}{r} dr.
\end{equation}
Since $R'$ is positive on $[0,a)$, the function $R$ itself is also
positive, $R(r)>0$ for $r\in (0,a]$. Due to the assumption
$P(0)>0$ the function $P^2(r)$ is not identically zero on this
segment. Therefore the integral at the right hand side of the
Eq.(\ref{eq:Ra12}) is strictly positive and hence $R'(a)\neq 0$,
what contradicts to the initial assumption.
\end{proof}
\begin{lem}
Consider two positive functions satisfying $P_1(r)>P_2(r)$ for
$r>0$. Then the difference $\Delta R\equiv R[q,P_1]-R[q,P_1]$ is
monotonously increasing function for $r>0$.
\end{lem}
\begin{proof}
Using the expansions near $r=0$ one can see  that the rate of
growth of $\Delta R$ near the origin is proportional to the value
of $P_{1}^2-P_{2}^2$ at the origin. Therefore there exists some
small value $r=a$  for which $\Delta R(a)>0$, $\Delta R'(a)>0$.
Again assuming that  at some point $r=b$ the derivative $\Delta
R'$ vanishes and integrating the equation for this quantity (which
is essentially the same as for $R'$ due to linearity of the Eq.
\ref{eq:Req}) we obtain
\begin{equation}\label{eq:Ra22}
    \Delta R'(b)=\Delta R'(a)\frac{a}{b}+\frac{1}{b}\int\limits_{a}^{b}\frac{\Delta R(P_{1}^2-P_{2}^2)}{r} dr.
\end{equation}
Similarly to the previous lemma, we obtain $\Delta R'(b)>0$ what
contradicts to the assumption. Thus $\Delta R'(r)$ is positive and
$\Delta R(r)$ is monotonously increasing for $r>0$.
\end{proof}

Now let us turn to the properties of the solution of the
Eq.(\ref{eq:Peq}) for $P$ with a given arbitrary regular function
$R(r)$.
\begin{lem}
If at some point $r=a$ one has  ${P(a)\geq 0}$, $P'(a)>0$, then
both $P(r)$ and $P'(r)$ are positive monotonously increasing
functions for $r>a$.
\end{lem}
\begin{proof}
Assume that there is some point $b>a$ such that  ${P'(a\leq
r<b)>0,}\: P'(b)=0$. Then $P$ will be a non-decreasing function on
the segment $[a,b]$ implying $P\geq 0$. Direct integration of the
Eq.(\ref{eq:Peq}) leads to
\begin{equation}\label{eq:Pa11}
P'(b)=P'(a)b/a+b\int\limits_{a}^{b}R^2P d\tilde{r}.
\end{equation}
The integrand at the right hand side  is non-negative, therefore
$P'(b)>0$ contradicting to the assumption. Thus  $P(r)$ is a
monotonously increasing function and from the Eq.(\ref{eq:Pa11})
it follows that $P'$ is   monotonously increasing  for $r>a$ too.
\end{proof}
Since the Eq.(\ref{eq:Peq}) is linear in $P$, it is invariant
under inversion $P\ra -P$, therefore if $P(a)\leq 0,\: P'(a)<0$
then $-P(r)$ and $-P'(r)$ are positive monotonously increasing
functions for $r>a$.

Now let us turn to the full system (\ref{eq:Req},~\ref{eq:Peq}).
One can easily see that there exist integral curves corresponding
to both cases considered above. Indeed, for small values of $|p|$
and large values of $q$ the derivative $P'$ with the growing of
$r$ soon becomes positive with $P$ also remaining positive. Vice
versa, for large $|p|$ and small $q$  the function $P$ soon
changes the sign while $P'$ also remains negative. In both cases
we will observe an infinite growth of $|P|$ as $r\ra\infty$.
Between these two sets of  $p,\,q$  there is a bifurcation set
such that the corresponding integral curve satisfies the
inequalities $P(r)>0$, $P'(r)<0$ as $r\ra\infty$. Then $|P|$ will
be a monotonously decreasing function at infinity. Thus we  will
obtain the heteroclinic trajectory  connecting two special points
of the system: the origin $(r=0,R=0)$ and infinity $(1/r=0,P=0)$.
Using the above lemmas one can prove that this solution is unique:
\begin{theor}
For each set of parameters ${q,k}$ there is a unique value of
parameter $p$ corresponding to the heteroclinic trajectory
connecting points $(r=0,R=0)$ and $(1/r=0,P=0)$.
\end{theor}
\begin{proof}
Let us assume that there are  two such trajectories. Since the
solution of the system is fully specified by the set of parameters
$(q,p,k)$, different trajectories must correspond to different
values of $p$, say $p_1>p_2$. The series expansion for $P$ near
the origin implies $\Delta P\equiv P_1-P_2\approx
(p_1-p_2)r^2/2>0$. This means that there is some point $r=a$,
where $\Delta P(a)>0,\:\Delta P'(a)>0$. Now consider the point
$r=b$ where $\Delta P'$ vanishes: $\Delta P'(a\leq r<b)>0$,
$\Delta P'(b)=0$. Since $\Delta P'$ is positive, the function
$\Delta P$ is monotonously increasing which implies that $P_1>P_2$
on the segment $[a,b]$ ( with both $P_1$, $P_2$ remaining
positive). Using the lemma 2, we find that $R_1>R_2$ on this
segment. Integration of the equation for $\Delta P$ gives
\begin{equation}
\Delta P'(b)=\Delta P'(a)b/a+b\int\limits_{a}^{b}(R_{1}^2
P_1-R_{2}^2 P_2) d\tilde{r}.
\end{equation}
The integrand is positive and the quantity $\Delta P'(a)$ is
positive too, therefore $\Delta P'(b)>0$ what contradicts to the
assumption. This means that $\Delta P(r)$ is monotonously growing
function for $r>a$ with the rate of growth not slower than $r^2$.
But as $r\ra\infty$, both $P_1$, $P_2$ vanish and $\Delta P$ has
to vanish too. Thus we arrive at  the contradiction again, hence
the assumption about the existence of several heteroclinic
trajectories is wrong. Therefore there is a unique value  $p=p_b$
for each set $(q,k)$ for which the corresponding solution
satisfies the boundary conditions $R(0)=0,\:P(0)=k\neq 0$ and
monotonously reaches its asymptotic form (\ref{eq:RPcondflat}).
\end{proof}
Thus the full solution space consists of the following sets:
\begin{itemize}
\item{Solutions with monotonously increasing $|P|$. Numerical
calculations show that these grow infinitely at some finite $r$,
i.e. they are singular.} \item{Solutions with monotonously
decreasing $|P|$. There is a unique bifurcation value  $p=p_b$ for
each set of $(q,k)$ for which such a solution can arise. These
solutions are globally regular.}
\end{itemize}
\par
Numerical integration of radial equations starting from the polar
axis $r=0$ shows that for any fixed value of $q$ both types of
solutions are realized. Singular solutions diverge at finite
distance $r=r_s$ as follows:
\begin{eqnarray}
&R(r)&\ra +\infty,\quad P(r)\ra +\infty,\quad {\mbox as} \quad
r\ra
 r_s,\\
&R(r)&\ra +\infty,\quad P(r)\ra -\infty, \quad{\mbox as}\quad r\ra
r_s.
\end{eqnarray}
\par
In the vicinity of the singularity, the system (\ref{eq:Req}),
(\ref{eq:Peq}) can be integrated analytically. In the leading
order one has
\begin{eqnarray}
R''=R\tilde{P}^2,\\
\tilde{P}''=R^2\tilde{P},
\end{eqnarray}
where $\tilde{P}=P/r_s$. This system is invariant under the
transformation $R\leftrightarrow \tilde{P}$ (actually this is the
symmetry (\ref{eq:gauget3})). Multiplying these equations on $R$
and  $\tilde{P}$ respectively one obtains
\begin{equation}RR''=\tilde{P}\tilde{P},''\end{equation} and thus there exist solutions
$\tilde{P}=\pm R$. Then the equation
\begin{equation}R''=R^3\end{equation} can be easily solved, and
find we the divergent terms as follows:
\begin{equation}
  R = \frac{\sqrt{2}}{r_s-r}, \qd P = \pm\frac{\sqrt{2}\;r_s}{r_s-r}.\\
\end{equation}
\par
Now discuss globally regular solutions.  It was confirmed
numerically that for any fixed value of $q$ there is  a unique
value $p=p_b$ such that for any $p>p_b$ and $p<p_b$ the integral
curves correspond to  singular solutions of the first and second
types respectively. As $p\ra p_b$ the singularity point $r_s$
monotonously moves away from the origin presumably to infinity. In
order to reconfirm the existence of regular solutions numerically
we have performed integration from infinity gluing it with the
solution started from the polar axis. Asymptotically the regular
solution to the system (\ref{eq:Req}--\ref{eq:Peq}) is given by
the following three-parameter family:
\begin{eqnarray}
 && R = (C_1\ln r+C_2)\alpha(r), \label{eq:Ra13}\\
 && P = D\sqrt{\frac{r}{\ln r}}\!\exp(-C_1r\ln
  r\!+\!(C_1\!-\!C_2)r)\beta(r),\quad
  \label{eq:Pa13}
\end{eqnarray}
where some functions $\alpha(r),\,\beta(r)$ satisfying the
boundary conditions $\alpha,\: \beta\ra 1$ as $r\ra\infty$. The
expansions for $\alpha, \beta$ are rather complicated because of
the presence of logarithmic terms $R$ and exponential terms for
$P$. In particular,
\begin{eqnarray}
    \beta(r) &=& 1-\frac{C_2}{2C_1\ln r}+o(\ln^{-1} r),\\
    \alpha(r) &=& 1+\left(\frac{P}{2C_1 r\ln r}\right)^2(1+o(\ln^{-1}
    r)).
\end{eqnarray}
Therefore any globally regular solution is fully specified by the
set parameters $(C_1,C_2,D)$ as well as by set $(q,p,k)$. In the
asymptotic set the parameter $D$ defines a scale, reflecting the
situation with the set $(q,p,k)$. Numerical integration provides
matching  of these two sets of parameters with each other.
\par
\begin{figure*}
\centering
\includegraphics{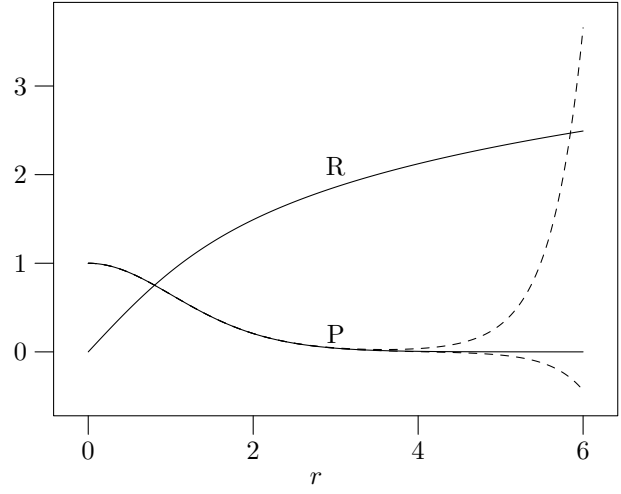}
\caption{Bifurcation of solutions in flat space. Dashed lines ---
the two singular solutions corresponding to the parameter values
$\:p=-0.9076$ and $\:p=-0.9078$ respectively.  Solid lines
correspond to the regular solution with the bifurcational value
$p=-0.90777863505\ldots$.}\label{fig:flat}
\end{figure*}
\begin{figure*}
\centering
\includegraphics{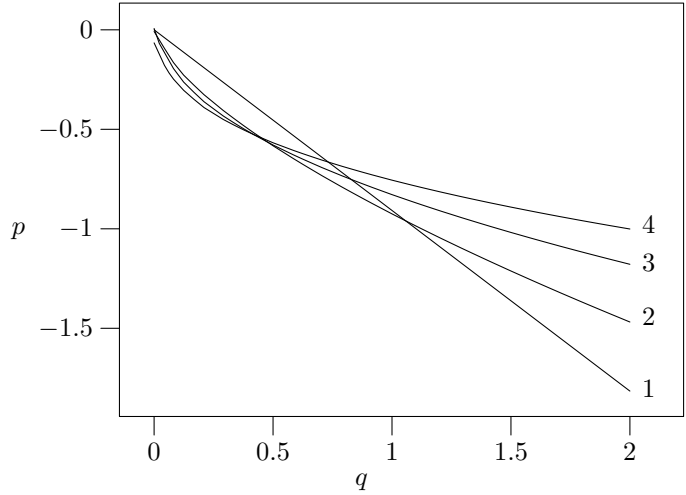}
\caption{Bifurcation curves for $k=1,2,3,4$ in the flat
space.}\label{fig:bifflat}
\end{figure*}

Previously we have obtained the necessary condition
(\ref{eq:Bcondflat}) of existence of regular asymptotically
vacuous solution.  Numerical integration confirms that for the
solutions at the bifurcation point $p=p_b$ this  condition is
satisfied indeed. Both regular and singular numerical solutions
for $q=1,\:k=1$ are shown in Fig.~\ref{fig:flat}. For other values
of $k$ the behavior of solutions is similar. Also, as we have
argued there are families of the solutions related by the scale
transformation. Numerical computations confirms that all solutions
with
\begin{equation}\label{eq:Bcond11}
p=p_b q^{\frac{2}{k+1}}
\end{equation}
are regular. They correspond to the  bifurcation curves of
 in the parameter space shown in Fig.~\ref{fig:bifflat}.
\subsection{Self-gravitating case}
As was argued before, the behavior of the full self-gravitating
solutions corresponding to small values of parameters  $q\ra 0,\:
p\ra 0$ must be the same as in the flat space. This implies the
existence of both regular and singular solutions in the
gravitating case too. Numerical integration of the   system
(\ref{eq:Rcurved}--\ref{eq:Ncurved}) proceeds along the same lines
with the boundary conditions obtained from the expansions on the
polar axis with free parameters $(q,p,k)$ (Tab. II). The main
difference is that now the asymptotic behavior of the metric
functions and $R$ at infinity is power-low with the linear
combinations of the Kasner parameters $(a,b,c)$ in the exponent.
For different sets of these parameters the linear combinations
with the same coefficients can be positive or negative which mean
either infinite growth or vanishing of the corresponding terms in
the expansion. For this reason the asymptotic solutions will be
different for different sets of $(a,b,c)$. We give here the
asymptotic expansions in the most symmetric case of the  values of
parameters: $a=2/3,b=2/3,c=-1/3$. In this case the functions
$(N,L,K,R)$ expand in powers of $x=r^{-1/3}$:
\begin{eqnarray}
  R &=& C_1+C_2x^2+\frac{24C_{2}^3}{\omega^2\gamma^2}\frac{x^4}{4!}+O(x^5), \label{eq:Ra14}\\
  N &=& \frac{\omega}{x^2}-\frac{C_{2}^2}{\omega\gamma^2}+\frac{\omega C_3}{\gamma}x+
  \frac{2C_{2}^4}{\omega^3\gamma^4}\frac{x^3}{3!}+O(x^5), \\
  L &=& N\frac{\gamma}{\omega}, \\
  K &=& \omega\gamma x-12\omega C_3\frac{x^4}{4!}+O(x^5).
\end{eqnarray}
The YM function $P$ has an exponential form:
\begin{equation}\label{eq:Pa14}
    P=\exp\left\{-\frac{3}{4\omega\gamma}\Lef\frac{C_1}{x^4}+\frac{2C_2}{x^2}-
    \frac{4C_{2}^3}{\omega^2\gamma^2}\ln{x}\Rig\right\}(D+O(x)).
\end{equation}

Altogether we have six parameters: $C_1,\, C_2$ and $D$ for the YM
functions as in the flat space, and four metric parameters:
$\omega,\; \gamma,\, C_3$, two of which can be changed by
rescalings of $t$ and $z$, and one independent Kasner parameter.
Together with three parameters on the axis $q,\:p,\:k$ one has ten
parameters to match the solutions obtained by the integration from
the axis and from the infinity. Since these solutions come from
the system of five second order differential equations the number
of parameters coincide with the number of matching conditions. The
numerical integration scheme is the same as in the flat case, and
the conservation of charges $Q_1,\,Q_2,\,H$ along the integral
curves is used to control the numerical error.

\begin{figure*}
\centering
\includegraphics{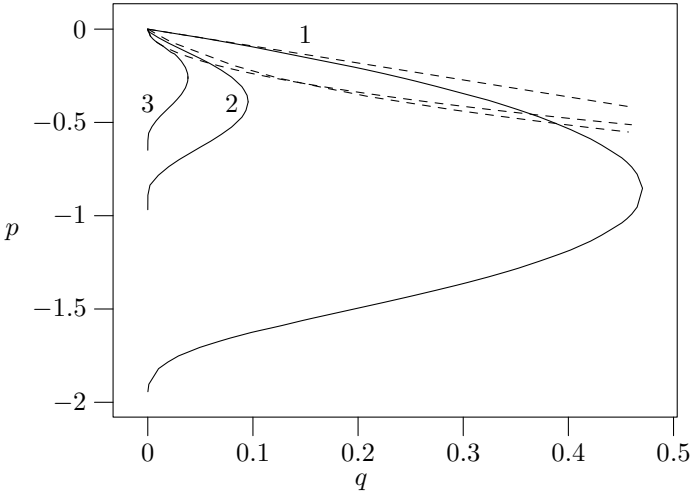}
\caption{Bifurcation curves for $k=1,2,3$ in the self-gravitating
case (solid lines) and in the flat space (dashed
lines).}\label{fig:bifcurv}
\end{figure*}

It turns out that in the gravitating case the bifurcation
parameters do not  satisfy the condition (\ref{eq:Bcond11})
anymore. The bifurcation curve has to be found numerically. Actual
calculations show that it is bounded within the finite region
 \begin{equation} 0\leq q\leq q_{max},\quad
0\geq p\geq p_{max},\end{equation} and for each $q$ in this region
there are {\em two} bifurcation parameters $p$. The shapes of the
bifurcations curves are similar for all $k$, but with  growing $k$
the region of parameters  becomes very narrow, this is illustrated
on Fig.~\ref{fig:bifcurv} for three lowest $k$. The condition
(\ref{eq:Bcond2}) is satisfied indeed for the regular solutions,
and  all other analytical results of the Sec. IV are also
confirmed by numerical computations.
\par
Using the Eqs. (\ref{eq:a}--\ref{eq:c}) we can relate the
asymptotic Kasner parameters to the value of the magnetic field on
the symmetry axis.  The weak field limit $\mathbb{B}\ra 0$
corresponds to the region  $q\ra 0, p\ra 0$ on the bifurcation
curve. In this case the metric approaches the Minkowski metric.
From the asymptotic relations
(\ref{eq:Ncurved2}--\ref{eq:NLKcurved2}) one can see that this
implies $M_\infty,\:W_\infty,\:X_\infty \ra 0,\;\Sigma_\infty\ra
1$, and the parameter $\beta$ in Eq.(\ref{eq:a}--\ref{eq:c})
approaches the value $1-\delta,\:\delta\ra 0$. In the first order
in $\delta$:
\begin{equation}
    N\sim r^\delta,\qd L\sim r,\qd K\sim
    r^{-\delta}.
\end{equation}
\par
As we move along the bifurcation curve, the field parameter
$\mathbb{B}$ grows up, the exponent of $L$ decreases, while the
exponent of $N$ increases. For  $\mathbb{B}=1/(2k)$ the metric
acquires an additional boost symmetry: from Eq.~(\ref{eq:beta}) it
follows that $\beta=0$ so the Kasner parameters are
$a=b=2/3,\:c=-1/3$. For this particular solution the metric
possesses  the asymptotical boost symmetry in the plane $(t,\vf)$
since $g_{tt}$ becomes proportional to $g_{\vf\vf}$  as
$r\ra\infty$.
\par
With $\mathbb{B}$ growing  further, the Kasner parameters satisfy
$a>b$. But still $g_{tt}$ and $g_{\vf\vf}$  grow at infinity while
$g_{zz}$ decreases. For the Melvin universe one has growing
$g_{zz}=g_{tt}$ and  decreasing $g_{\vf\vf}$. When the bifurcation
curve approaches the $p$-axis, which corresponds to the Abelian
case with $R=0$, the metric is of the Melvin type. It is
interesting to observe how  these different types of metric
transform into each other. It turns out that for $q\ra 0$ but
finite $p$ the  functions $R,\:R'$ are both small, while $P,\:P'$
are not. This is why only the $z$-string (associated with the
function $P$) persists in the vicinity of the symmetry axis. It
generates the Melvin metric with  growing $r$, while  the energy
density decreases as a power-law. In this region the stress-energy
tensor exhibits the Melvin symmetry:
\begin{equation}
    T_{0}^{0}=-T_{r}^r=-T_{\vf}^{\vf}=T_{z}^z.\label{eq:Zsim}
\end{equation}
But for larger $r$, the functions $R$ and $R'$ become
non-negligible, and the energy density of the $\vf-$string
increases and reaches its maximum Fig.~\ref{fig:curv2s2}. In this
region the energy density of the $z-$string decreases
exponentially. So  at the small distances from the axis the
$z-$string dominates and generates the Melvin metric, while at
larger distances  the $\vf$-string dominates,  transforming the
metric asymptotically into the form (\ref{eq:a}--\ref{eq:c}). In
this region the components of the stress-energy tensor satisfy the
relations
\begin{equation}
    T_{0}^{0}=-T_{r}^r=T_{\vf}^{\vf}=-T_{z}^z.\label{eq:Fsim}
\end{equation}

As $p\ra p_{max}$, the boundary between these two regimes  moves
to infinity. Since  strings cores are largely separated, one can
introduce the effective radius $r_0$ such that  the mass of the
$z-$-string $M_z$ is concentrated in the region $(0,~r_0)$, while
the mass of the $\vf-$string $M_\vf$ corresponds to the region
$(r_0,~\infty)$. We define then
\begin{equation}
    M_z=M(r_0),\: M_\vf=M_\infty-M(r_0),
\end{equation}
and similarly for the quantities $W,\:X,\:\Sigma$. The solution
within the region $(0,r_0)$ can be considered as  Melvin, so  it
is known that $M_z=2$, and due to the symmetry (\ref{eq:Zsim}) one
has $M_z=-W_z=X_z=2$, $\Sigma_z=3$. The integration of the Eqs.
(\ref{eq:N1}--\ref{eq:K1}) gives:
\begin{eqnarray}
  N'L K &\ra& M_\vf+M_z=M_\vf+2, \\
  N L'K &\ra& W_\vf+W_z+1=W_\vf-1, \\
  N L K' &\ra& X_\vf+X_z=X_\vf+2.
\end{eqnarray}
Using the asymptotic symmetry (\ref{eq:Fsim}), one finds
$M_\vf=W_\vf=-X_\vf$ and the Kasner parameters can be expressed in
terms of the $\vf-$string mass $M_\vf$ only:
\begin{equation}
    a=\frac{M_\vf+2}{M_\vf+3},\qd b=\frac{M_\vf-1}{M_\vf+3},\qd
    c=\frac{2-M_\vf}{M_\vf+3}.
\end{equation}
For them the first Kasner condition is satisfied automatically,
and the second gives the following equation for $M_\vf$:
\begin{equation}
    M_\vf(M_\vf-4)=0.
\end{equation}
So we have two possibilities: either $M_\vf=0$ implying the pure
Melvin metrics, or $M_\vf=4$ (in our dimensionless units) which
gives the limiting solution $a=6/7,\:b=3/7,\:c=-2/7$.  For small
values of the parameter $q\ra 0$ when the above picture of well
separated strings is valid, the mass of the $\vf-$string
approaches the value $4$. Recall that the pure Melvin case,  the
mass of the $z-$string approaches the value $2$ as $p\ra 0$.

\begin{figure*}
\centering
\includegraphics{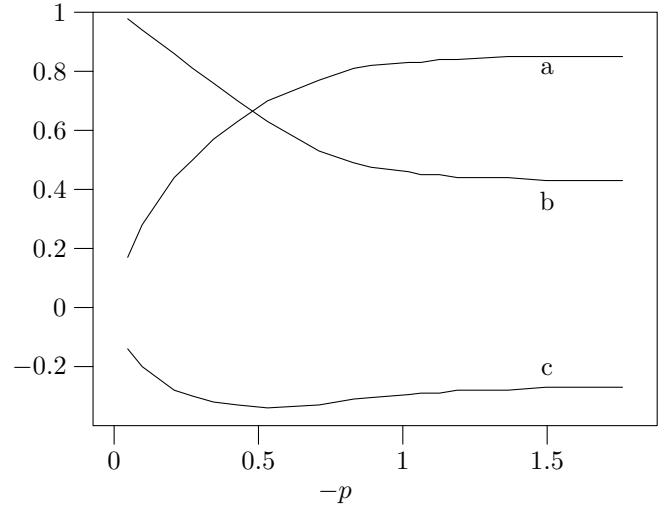}
\caption{Kasner asymptotic parameters corresponding to the
solutions with $k=1$.} \label{fig:all}
\end{figure*}

\begin{figure*}
\centering
\includegraphics{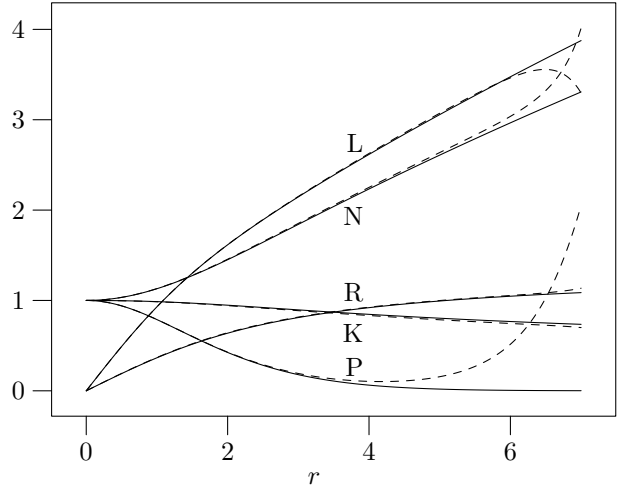}
\caption{ The metric and field functions for the self-gravitating
solutions.  Dashed lines
--- the singular solution $(q=0.39,\:p=-0.5)$, solid lines --- the
regular solution
$(q=0.3862649505\ldots,\:p=-0.5)$.}\label{fig:curved}
\end{figure*}

\begin{figure*}
\centering
\includegraphics{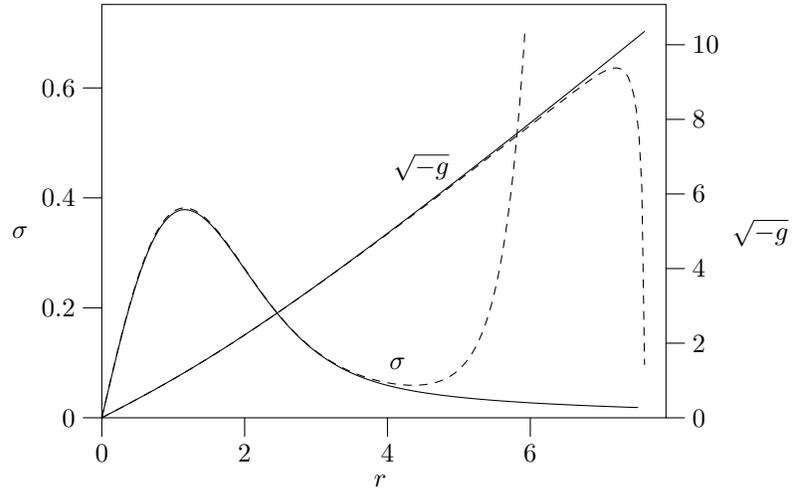}
\caption{ The square root of the metric tensor $\sqrt{-g}$ and the
energy density $\sigma=T_{0}^0\sqrt{-g}$ for self-gravitating
solutions. Dashed lines
--- the singular solution $(q=0.39,\:p=-0.5)$, solid lines --- the
regular solution at bifurcation
$(q=0.3862649505\ldots,\:p=-0.5)$.}\label{fig:curved2}
\end{figure*}

\begin{figure*}
\centering
\includegraphics{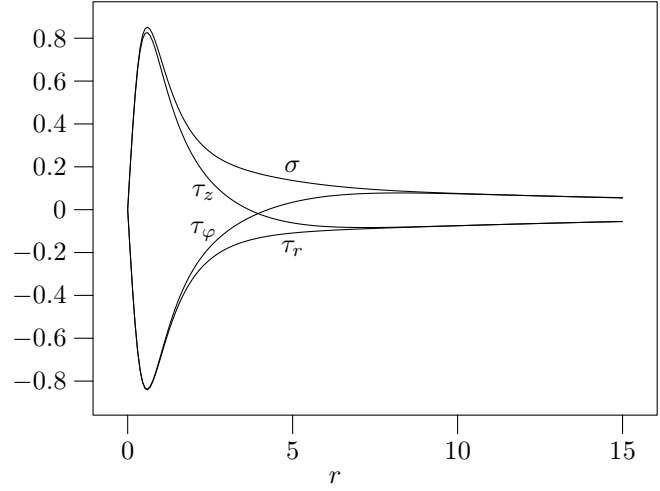}
\caption{ The components of the stress-energy tensor
$\sigma=T^{0}_{0}\sqrt{-g}$, $\tau_i=T^{i}_{i}\sqrt{-g}$ for the
self-gravitating configuration.} \label{fig:curp}
\end{figure*}

The numerical results for $k=1$ are presented on
 Fig.~\ref{fig:all}, where  the dependence of Kasner parameters
on the field strength $p$ is shown. A typical numerical solution
is shown on  Figs.~\ref{fig:curved},\ref{fig:curved2}. The
 Fig.\ref{fig:curp}  describes how the components of the
stress-energy tensor switch from the symmetry (\ref{eq:Zsim}) to
 (\ref{eq:Fsim})  with  growing  $r$. On
 Fig.~\ref{fig:curv2s}  one can see the core of the $z-$string in
the vicinity of the axis which form\textbf{}s the Melvin metrics.
 Fig.~\ref{fig:curv2s2}  demonstrates that at some distance from
the axis there is another pick of the energy density --- the core
of the $\vf-$string, which changes the signs in the exponents of
$g_{zz}$ and $g_{\vf\vf}$.

\par
Since at $p=p_{max}$ both Abelian and non-Abelian solutions
coincide, one can connect the corresponding parameters. The
asymptotical value of the potential $A_\phi$ for the Melvin
solution is $2/\mathbb{B}$. In the non-Abelian case we have shown
that the corresponding asymptotic value is $-k$. This means that
\begin{equation}\label{eq:Bmax}
    \mathbb{B}_{max}=-p_{max}=2/k.
\end{equation}
Numerical computations confirm exactly this result, as illustrated
on Fig.~\ref{fig:bifcurv}. Furthermore, in the Abelian case the
bifurcation condition (\ref{eq:Bcond20})
\begin{equation}
    2=k\mathbb{B}_{max},
\end{equation}
gives again (\ref{eq:Bmax}). In this limit the following
inequality holds $\Sigma_\infty\geq 3$ (the strict equality holds
for the Melvin solution) and thus $\beta\geq -1$. Therefore the
parameter $\beta$ in Eq.(\ref{eq:a}--\ref{eq:c}) takes values
within the interval $[-1,1]$, which ensures positivity of the
expression under the square root. But the signs before the square
root are different in the Abelian and non-Abelian cases.
\par
\begin{figure*}
\centering
\includegraphics{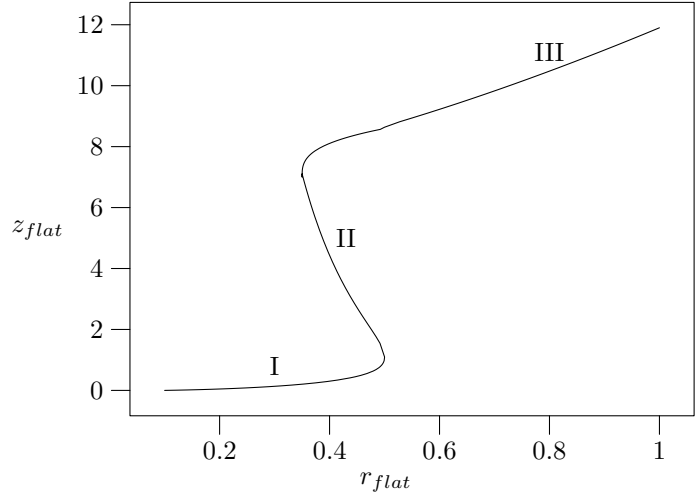}
\caption{The geometry of the section $z=\rm const.$ of the
solution carrying the metric $dl^2=dr^2+L^2(r)d\varphi^2$ as
embedded into the three-dimensional flat space (the desired
surface is the figure of rotation of this curve around the
vertical axis). The coordinate along the horizontal axis is
$r_{flat}=\sqrt g_{\vf\vf}$, the vertical coordinate is defined by
$dz_{flat}^2=dr^2-dr_{flat}^2$. Three regions correspond to the
nearly flat metric, and that of the Melvin straight and circular
strings respectively.}\label{fig:L}
\end{figure*}
\begin{figure*}
\centering
\includegraphics{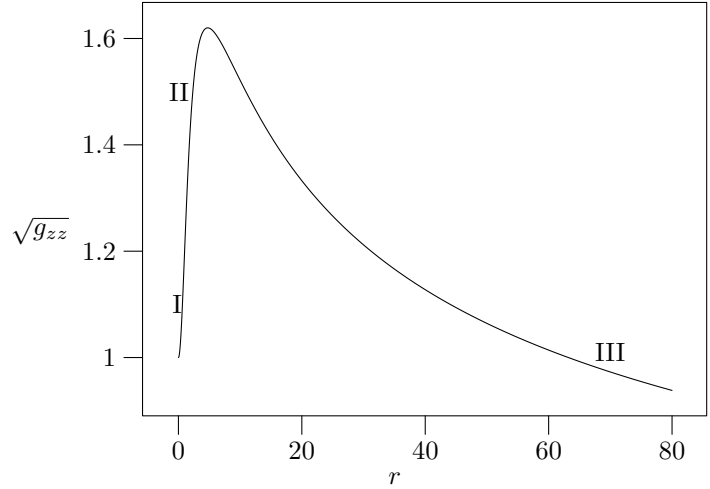}
\caption{Plot of the metric function $K(r)$ showing an effective
longitudinal thickness of the configuration  along the $z-$axis as
a function of the radial. Three regions are shown which correspond
to the nearly flat metric, the straight string core and the
circular string.}\label{fig:K}
\end{figure*}

The space generated by the $\vf$-string can be visualized as the
surface of a cup. Since $0<b<1$, the circumference of the circle
bounding the disk of the radius $r$ centered on the axis and
perpendicular to it is growing up slower than $r$. If we embed the
surface $(r,\vf)$ carrying the corresponding metric into the flat
space of one dimension more, it will look like a cup whose shape
looks asymptotically as a figure of rotation of the curve
$z_{flat}=r_{flat}^{1/b}$ Fig.~\ref{fig:L}. Two such surfaces
lying at the unit distance along the axis will form a thick disk
with the thickness decreasing asymptotically as
$z_{flat}=r_{flat}^{-|c|}$, see Fig.~\ref{fig:K}.
\par
Solutions corresponding to any set of parameters $(q,p,k)$  not
belonging to the bifurcation curve are singular. Their behavior is
similar to that in the flat case, but now there is a singularity
of the metric, not of the gauge field. At the singularity the
determinant of the metric vanishes, $g(r_s)=0$, while the
Kretschmann scalar (\ref{eq:krechman}) diverges as $(r_s-r)^{-4}$.
The corresponding expansions  of the metric and the field
functions in terms of the deviation $x=r_s-r$ can be found
analytically. It turns out that the expansion of field functions
contains only the integer exponents while that of the metric
functions --- only half-integer exponents. In the leading order in
$x$
\begin{equation}
    P,\:R\ra {\rm const},\; N\ra x^{-1/2},\; L,\:K\ra x^{1/2}.
\end{equation}
An example of such an expansion for $k=1$ is given in
Tab.~\ref{tb:curvedsing}.

At the end of this section we would like to mention that the
obtained results for the bifurcation conditions
(\ref{eq:Bcond20},~\ref{eq:Bcondflat}), as well as the asymptotic
expansions
(\ref{eq:Ra13}-\ref{eq:Pa13},~\ref{eq:Ra14}-\ref{eq:Pa14}) does
not depend on the assumed discrete nature of $k$. The asymptotic
behavior of the system will be the same for non-integer values of
$k$.

\section{Conclusion}
In this paper, we have continued investigation of static
cylindrically symmetric purely magnetic SU(2) EYM configurations
started in \cite{GDV06}. We extended analysis of bifurcation
manifolds, gave a (non-rigorous) proof of existence, considered in
more detail structure of solution space near the high-curvature
limit and described in detail singular solutions. Our set of the
integral sum rules   allows one to relate the asymptotic
parameters of the solution to the boundary data on the polar axis
and to simplify the numerical integration.  The values of the
Kasner exponents can be found analytically in some cases.
Otherwise they are obtained numerically starting with the data on
the symmetry axis parameterized by the value of the magnetic
field.
\par
The geometric structure and the matter distribution for soliton
solutions obtained suggest an interpretation of the solution as
describing a pair of straight and circular magnetic strings.
Asymptotically the metric exhibits the boost symmetry in the plane
$(t,\vf)$ similar to the $(t,z)$ boost symmetry of the usual
Melvin solution.
\par
Some generalizations can be suggested, in particular, similar
structure should exist in the bosonic sector of the
Freedman--Shwartz SU(2)$\times$SU(2) gauged supergravity. An
intriguing question is whether the supersymmetric solutions may
exist. The corresponding Abelian U(1)$\times$U(1) configurations
were considered in \cite{Radu:2003av} showing the absence of
supersymmetry for the Melvin-type configuration. But as was shown
in \cite{Chamseddine:1997mc, Gubser:2001eg},  spherically
symmetric static pure magnetic configurations admits supersymmetry
of gauged supergravity only in the case of truncation to
SU(2)$\times$U(1). So  supersymmetry for cylindrically symmetric
gauge field configurations is still not excluded in the
non-Abelian sector.
\par
As other directions we could mention an inclusion of the electric
component of the YM field, as well as dressing the solutions with
additional structure such as propagating gravitational waves
\cite{Garfinkle:1992dz} and black holes  \cite{Hiscock:1980zf,
Ortaggio:2003ri}, similar to that found for the Melvin background.
\begin{acknowledgments}
   The work was supported in part by
the RFBR grant 02-04-16949.
\end{acknowledgments}

\newpage
\extrarowheight=10pt
\begin{table*} \caption{Expansions at the origin in flat space}
\label{tb:flat}
\begin{tabular}{|c|l|}
  \hline
  $k$ & \multicolumn{1}{c|}{$R$}\\
  \hline
  1 & $\displaystyle{q r+ \frac{3}{4}pq\frac{r^3}{3!}+\left(\frac{15}{8}p^2q+\frac{5}{4}q^3\right) \frac{r^5}{5!}+
        \left(\frac{315}{64}p^3q+\frac{385}{16}pq^3\right) \frac{r^7}{7!}
        +O(r^{9})}
        $ \\
  2 &
  $\displaystyle{q\frac{r^2}{2!}+\!2pq\frac{r^4}{4!}+\!\frac{105}{16}p^2q\frac{r^6}{6!}+\!\left(\frac{105}{4}p^3q+\!28q^3\right)\!
  \frac{r^8}{8!}+\!\left(\frac{17325}{128}p^4q+\!\frac{9345}{8}pq^3\right)\!\frac{r^{10}}{10!}+\!O(r^{12})}$ \\
  3 & $\displaystyle{q\frac{r^3}{3!}+\frac{15}{4}pq\frac{r^5}{5!}+\frac{273}{16}p^2q\frac{r^7}{7!}+\frac{1449}{16}p^3q\frac{r^9}{9!}+
  \left(\frac{72765}{128}p^4q+\frac{2475}{4}q^3\right)\frac{r^{11}}{11!}+O(r^{13})}$ \\
  4 & $\displaystyle{q\frac{r^4}{4!}+6pq\frac{r^6}{6!}+\frac{147}{4}p^2q\frac{r^8}{8!}+\frac{495}{2}p^3q\frac{r^{10}}{10!}
  +\frac{479655}{256}p^4q\frac{r^{12}}{12!}+O(r^{14})}$\\[7pt]
  \hline
   & \multicolumn{1}{c|}{$P$} \\
  \hline
  1 & $\displaystyle{1+p\frac{r^2}{2!}+3q^2\frac{r^4}{4!}+\frac{45}{2}pq^2\frac{r^6}{6!}+\left(\frac{1155}{8}p^2q^2+\frac{245}{2}q^4
  \right)\frac{r^{8}}{8!}+O(r^{10})}$ \\
  2 & $\displaystyle{2+p\frac{r^2}{2!}+15q^2\frac{r^6}{6!}+245pq^2\frac{r^8}{8!}+\frac{26775}{8}p^2q^2\frac{r^{10}}{10!}+
  \left(\frac{197505}{4}p^3q^2+26334q^4\right)\frac{r^{12}}{12!}+O(r^{14})}$ \\
  3 & $\displaystyle{3+p\frac{r^2}{2!}+70q^2\frac{r^8}{8!}+\frac{4095}{2}pq^2\frac{r^{10}}{10!}+\frac{367983}{8}p^2q^2\frac{r^{12}}
  {12!}+\frac{16081065}{16}p^3q^2\frac{r^{14}}{14!}+O(r^{16})}$ \\
  4 & $\displaystyle{4+p\frac{r^2}{2!}+315q^2\frac{r^{10}}{10!}+14553pq^2\frac{r^{12}}{12!}+\frac{963963}{2}p^2q^2\frac{r^{14}}{14!}+
  \frac{58359015}{4}p^3q^2\frac{r^{16}}{16!}+O(r^{18})}$ \\[7pt]
  \hline
\end{tabular}
\end{table*}

\begin{table*}
\caption{Expansion at the origin in curved space ($k=1$)}
\label{tb:curved}
\begin{tabular}{|c|l|}
   \hline
  R & $\displaystyle{q r + \Lef\frac{3}{4}pq+p^2q-\frac{3}{2}q^3\Rig\frac{r^3}{3!}+O(r^{5})}$ \\
  P & $\displaystyle{1+p\frac{r^2}{2}+\Lef 3q^2-5p^3\Rig\frac{r^4}{4!}+O(r^6)}$ \\
  N & $\displaystyle{1\!+\!\!\Lef\!\frac{1}{2}p^2\!+\!q^2\!\Rig\!\!\frac{r^2}{2!}\!+\!\Lef \!4p^2q^2\!-\!p^4\!+\!\!\frac{9}{2}pq^2\!+\!3q^4\!\Rig\!\!\frac{r^4}{4!}
  +\!O(r^6)}$ \\
  L & $\displaystyle{r-2p^2\frac{r^3}{3!}+\Lef\frac{35}{2}p^4-12pq^2-6q^4\Rig\frac{r^5}{5!}+O(r^7)}$ \\
  K & $\displaystyle{1\!+\!\!\Lef\!\frac{1}{2}p^2\!-\!q^2\!\Rig\!\!\frac{r^2}{2!}\!+\!\Lef \!3q^4\!-\!4p^2q^2\!-\!p^4\!-
  \!\frac{9}{2}pq^2\!\Rig\!\!\frac{r^4}{4!}+\!O(r^6)}$\\[7pt]
  \hline
\end{tabular}
\end{table*}

\begin{table*}
\caption{Expansion at the singular point in curved space ($k=1$)}
\label{tb:curvedsing}
\begin{tabular}{|c|l|}
   \hline
  R & $\displaystyle{R_0-\frac{K_1^2}{2R_0}x+\frac{5L_1K_{1}^4+2K_{1}^3P_2R_0-20R_{0}^4L_1}{4L_1R_{0}^2}\frac{x^2}{2!}}+O(x^{3})$ \\
  P & $\displaystyle{\frac{L_1K_1}{2R_0}-\frac{L_1R_0}{K_1}x+P_2\frac{x^2}{2!}+O(x^{3})}$ \\
  N & $\displaystyle{N_0x^{-1/2}+\frac{N_0(3L_1K_{1}^4+2R_{0}^4L_1+K_{1}^3P_2R_0)}{4L_1R_{0}^2K_{1}^2}x^{1/2}+O(x^{3/2})}$ \\
  L & $\displaystyle{L_1x^{1/2}+\frac{L_1K_{1}^4-2R_{0}^4L_1-K_{1}^3P_2R_0}{4R_{0}^2K_{1}^2}x^{3/2}+O(x^{5/2})}$ \\
  K & $\displaystyle{K_1x^{1/2}+\frac{3L_1K_{1}^4-14R_{0}^4L_1+K_{1}^3P_2R_0}{4R_{0}^2L_1K_{1}^1}x^{3/2}+O(x^{5/2})}$\\[7pt]
  \hline
\end{tabular}
\end{table*}

\newpage

\begin{figure*}
\centering
\includegraphics{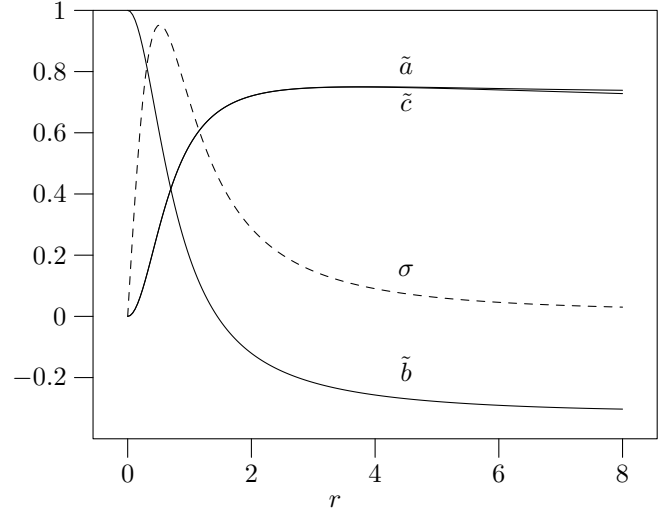}
\caption{The metric of the regular solution with $p=-1.85$. Solid
lines correspond to the effective exponents of metric functions
$\tilde{a}_i=(\ln{N_i})'r$ where $i$ labels $a,\:b,\:c$ and
$N,\:L,\:K$ respectively. When $\tilde{a}_i\approx \rm const$, the
metric functions $N_i\sim r^{\tilde{a}_i}$. Dashed line --- the
energy density $\sigma=T_{0}^0\sqrt{-g}$.}\label{fig:curv2s}
\end{figure*}

\begin{figure*}
\centering
\includegraphics{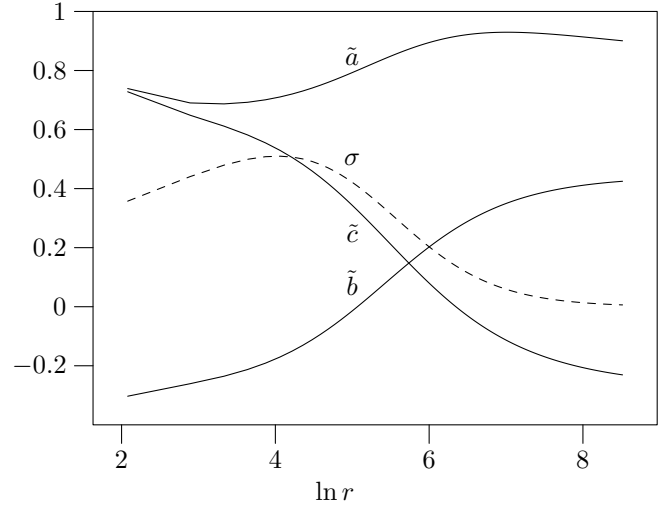}
\caption{Continuation of the previous plot Fig.~\ref{fig:curv2s}
with an enhanced (factor of 100) scale for
$\sigma$.}\label{fig:curv2s2}
\end{figure*}

\end{document}